\documentclass[]{emulateapj}
\usepackage{apjfonts}
\usepackage{natbib}
\usepackage{graphicx}
\usepackage{epstopdf}
\usepackage{multirow} 
\usepackage{longtable,lscape} 
\usepackage{color}
\usepackage{xspace}
\usepackage{xargs}
\usepackage{url}
\usepackage{gensymb}

\newcommand{\HST}{\emph{HST}}

\newcommand{\Ha}{H$\alpha$\xspace}
\newcommand{\ma}{MACS0717\xspace}
\newcommand{\mb}{MACS1423\xspace}

\shorttitle{Spatially Resolved SFR in cluster galaxies}
\shortauthors{Vulcani et al.}

\begin{document}
\title{THE GRISM LENS-AMPLIFIED SURVEY FROM SPACE (GLASS). V. Extent and spatial distribution of star formation in $z\approx0.5$ cluster galaxies} 

\author{Benedetta Vulcani\altaffilmark{1}}
\author{Tommaso Treu\altaffilmark{2}}
\author{Kasper B. Schmidt\altaffilmark{3}}
\author{Bianca M. Poggianti\altaffilmark{4}}
\author{Alan Dressler\altaffilmark{5}}
\author{Adriano Fontana\altaffilmark{6}}
\author{Marusa Brada\v{c}\altaffilmark{7}}
\author{Gabriel B. Brammer\altaffilmark{8}}
\author{Austin Hoag\altaffilmark{7}}
\author{Kuan-Han Huang\altaffilmark{7}}
\author{Matthew Malkan\altaffilmark{2}}
\author{Laura Pentericci\altaffilmark{6}}
\author{Michele Trenti\altaffilmark{9}}
\author{Anja von der Linden\altaffilmark{10,11}}
\author{Louis Abramson\altaffilmark{2}}
\author{Julie He\altaffilmark{7}}
\author{Glenn Morris\altaffilmark{11,12}}

\affil{\altaffilmark{1}Kavli Institute for the Physics and Mathematics of the Universe (WPI), The University of Tokyo Institutes for Advanced Study (UTIAS), the University of Tokyo, Kashiwa, 277-8582, Japan}
\affil{\altaffilmark{2}Department of Physics and Astronomy, University of California, Los
Angeles, CA, USA 90095-1547}
\affil{\altaffilmark{3}Department of Physics, University of California, Santa Barbara, CA, 93106-9530, USA}
\affil{\altaffilmark{4}INAF-Astronomical Observatory of Padova, Italy}
\affil{\altaffilmark{5}The Observatories of the Carnegie Institution for Science, 813 Santa Barbara St., Pasadena, CA 91101, USA}
\affil{\altaffilmark{6}INAF - Osservatorio Astronomico di Roma Via Frascati 33 - 00040 Monte Porzio Catone, I}
\affil{\altaffilmark{7}Department of Physics, University of California, Davis, CA, 95616, USA}
\affil{\altaffilmark{8}Space Telescope Science Institute, 3700 San Martin Drive, Baltimore, MD, 21218, USA}
\affil{\altaffilmark{9}School of Physics, University of Melbourne, VIC 3010, Australia}
\affil{\altaffilmark{10}Dark Cosmology Centre, Niels Bohr Institute, University of Copenhagen Juliane Maries Vej 30, 2100 Copenhagen {\O}, DK}
\affil{\altaffilmark{11}Kavli Institute for Particle Astrophysics and Cosmology, Stanford University, 452 Lomita Mall, Stanford, CA  94305-4085, USA}
\affil{\altaffilmark{11}SLAC National Accelerator Laboratory, 2575 Sand Hill Road, Menlo Park,    CA 94025, USA}

\begin{abstract}
We present the first study of the spatial distribution of star
formation in $z\sim0.5$ cluster galaxies.  The analysis is based on
data taken with the Wide Field Camera 3 as part of the Grism
Lens-Amplified Survey from Space (GLASS).  We illustrate the
methodology by focusing on two clusters (MACS0717.5+3745 and
MACS1423.8+2404) with different morphologies (one relaxed and one
merging) and use foreground and background galaxies as field control
sample. The cluster+field sample consists of 42 galaxies with stellar
masses in the range 10$^8$-10$^{11}$ M$_\odot$, and star formation
rates in the range 1-20 M$_\odot \, yr^{-1}$. Both in clusters and in
the field, \Ha\ is more extended than the  rest-frame UV continuum in 60\% of the cases,
consistent with diffuse star formation and inside out growth. 
 In $\sim$ 20\% of the cases, the  \Ha\
emission appears more extended in cluster galaxies than in the field,
pointing perhaps to ionized gas being stripped and/or star formation
being enhanced at large radii.   The peak of the
\Ha\ emission and that of the continuum are offset by
less than 1 kpc. We investigate trends with  the hot gas density
as traced by the X-ray emission, and with the surface mass density as
inferred from gravitational lens models and find no conclusive results. The diversity of
morphologies and sizes observed in \Ha\ illustrates the complexity of
the environmental process that regulate star formation. Upcoming
analysis of the full GLASS dataset will increase our sample size by
almost an order of magnitude, verifying and strengthening the
inference from this initial dataset.

\end{abstract}


\section{Introduction}

Over the past decade, observations have shown that the star formation
activity in galaxies has strongly declined since $z\sim 2$ \citep[see,
e.g.,][]{hopkins06, madau14}, with a large number of star-forming
galaxies evolving into passive galaxies at later times, and the star
formation rate at fixed mass progressively decreasing \citep{bell04,
bell07, noeske07, daddi07, karim11}.  The evolution of the star
formation activity is coupled to the evolution of galaxy morphologies
\citep{poggianti09}, with a significant fraction of today's early-type
galaxies having evolved from late types at relatively recent epochs.
Even though transformations occur both in galaxy clusters
\citep{dressler97, fasano00} and in the field \citep{oesch10,
capak07}, the strength of the decline has been found to depend on 
environment: galaxies in clusters experience a stronger evolution in
star formation activity compared to galaxies in the field
\citep[e.g.,][]{poggianti06, cooper06, guglielmo15}.

Central for our progress in understanding galaxy evolution is
identifying the cause of the decline of star formation and of the
emergence of the different galaxy types. The mass of galaxies and the
environment where they reside are generally believed to play a role
for quenching the star formation \citep[e.g.,][]{peng10}, but the
specific physical mechanisms involved  remain obscure.
There is no consensus on whether there is one process that dominates
quenching across all environments or whether some processes play a larger role
in driving galaxy evolution in dense environments than they do in the
field \citep{butcher84, poggianti99, dressler99, treu03, dressler13}.

Each of the processes that have been proposed to quench star formation
in galaxies should leave a different signature on the spatial
distribution of the star formation activity within the galaxy.  For
example, ram-pressure stripping from the disk due to the interaction
between the galaxy interstellar medium (ISM) and the intergalactic
medium \citep[IGM,][]{gunngott72} is expected to partially or
completely remove the ISM, leaving a recognizable pattern of star
formation with truncated \Ha disks smaller than the undisturbed
stellar disk \citep[e.g.,][]{yagi15}. Strangulation, which is the
removal of the hot gas halo surrounding the galaxy either via
ram-pressure or via tidal stripping by the halo potential
\citep{larson80, balogh00}, should deprive the galaxy of its gas
reservoir, and leave the existing interstellar medium in the disk to
be consumed by star formation.  Strong tidal interactions and mergers,
tidal effects of the cluster as a whole and harassment, that is the
cumulative effect of several weak and fast tidal encounters
\citep{moore96}, thermal evaporation \citep{cowie77} and
turbulent/viscous stripping \citep{nulsen82} can also deplete the gas
in a non homogeneous way.

In order to address how star formation is suppressed in the different
regions of the galaxy, a key ingredient is the spatial distributions
of both the past star formation, as traced by the existing stellar population,  
and of the instantaneous star
formation. The latter can be traced by the \Ha line
emission as it scales with the quantity of ionizing photons produced
by hot young stars \citep{kennicutt98}.

In the local universe, a few studies have focused on the analysis of
\Ha spatial distribution of a limited number of systems in clusters
\citep[e.g.][]{merluzzi13,fumagalli14}, detecting debris of material
that appear to be stripped from the main body of the galaxy, and whose
morphology is suggestive of gas-only removal mechanisms, such as ram
pressure stripping.

However, our current understanding is that much of the activity in cluster galaxies happens
beyond the local universe at $z=0-1$ and it is therefore essential to
gather information in this redshift range. A number of \Ha surveys up
to $z\sim$1 have been undertaken in the field using narrow-band
imaging
\citep[e.g.,][]{sobral13} and with WFC3 grism observations
\citep[e.g.,][]{atek10, straughn11}. In clusters, narrow-band \Ha\
studies are available for just a few systems at $z = 0.3-1$
\citep{kodama04, finn05, koyama11} and a few other higher-$z$ overdense
regions (\citeauthor{kurk04a} 2004a; \citeauthor{kurk05} 2004b; \citealt{geach08, hatch11, koyama13}). These ground-based studies provide
integrated \Ha fluxes, and no spatial  information.

Recently, spatially resolved star formation maps at $z\sim1$ have been
obtained for field galaxies using both the ACS I band and the G141 grism on the Wide Field
Camera 3 (WFC3) on board the Hubble Space Telescope (HST) as part of
the 3D-HST Survey (\citealt{vandokkum11, brammer12, schmidt13}; Momcheva et al. in prep).
\citet{nelson12,
nelson13} mapped the \Ha and stellar continuum with high resolution
for $\sim$ 60 galaxies and showed that star formation broadly
follows the rest-frame optical light, but is slightly more extended.
By stacking the \Ha emission, they measured structural parameters of
stellar continuum emission and star formation, finding that star
formation occurred in approximately exponential distributions.  They
concluded that star formation at
$z\sim$1 generally occurred in disks. 

 \cite{wuyts13}  expanded the sample analyzed by \citet{nelson12,
nelson13} and characterized   the resolved stellar populations of $\sim$500 massive star-forming galaxies, 
with multi-wavelength broad-band imaging from CANDELS \citep{wuyts12} and \Ha surface brightness profiles. 
They found the \Ha morphologies to resemble more closely those observed in the 
ACS I band than in the WFC3 H band, especially for the larger systems. They also found that  the rate of ongoing star formation per unit area tracks the amount of stellar mass assembled over the same area.  Off-center clumps are characterized by enhanced \Ha equivalent widths, bluer broad-band colors and correspondingly higher specific star formation rates (SFRs) than the underlying disk, implying they are a star formation phenomenon.

 More recently, \cite{nelson15},
exploiting a much larger sample, studied the behavior of the \Ha
profiles above and below the main sequence and showed that star
formation is enhanced at all radii above the main sequence, and
suppressed at all radii below the main sequence.

In this paper we present a pilot study characterizing  the spatial
distribution of the \Ha emission in cluster galaxies beyond the local
universe based on WFC3-IR data drawn from the Grism Lens-Amplified
Survey from Space (GLASS; GO-13459; PI: Treu,\footnote{\url{http://glass.physics.ucsb.edu} } \citealt{schmidt14, treu15}).
The GLASS G102 data yield spatially resolved \Ha fluxes for
all star-forming  galaxies in the core ($<1$ Mpc) of 10 clusters
at $z = 0.31 - 0.69$, with an order of magnitude improvement in sensitivity compared to
 previous studies \citep{sobral13}. 
Each cluster is observed at two different position angles. These two orientations
allow us to mitigate the impact of contamination from overlapping spectra, and reliably
measure for the first time the relative position of the \Ha\ emission with respect to the
 continuum. 

We illustrate the methodology and first results of
this approach by analyzing two of the ten clusters in the GLASS
sample. Among the first clusters that have been observed by GLASS we
selected MACS0717.5+3745 (hereafter \ma) and MACS1423.8+2404  (hereafter \mb) based on the following
criteria. First, we required similar redshift, so as to minimize
evolutionary effects and differences in the sensitivity/selection
function ($z=0.55$). Second, we selected them to be in very different
dynamical states (MACS1423 is relaxed, MACS0717 is an active merger),
so as to span the range of expected environments. A homogeneous
control field sample is obtained by selecting objects in the immediate
foreground and background of the two clusters. In total the sample
presented in this pilot paper consists of 42 objects, evenly
distributed between the two clusters and the field (15 and 10 cluster galaxies and 9 and 8 field 
galaxies in \ma and \mb, respectively). 
In a forthcoming
paper, to appear after the complete GLASS data  have been
processed, we will present an analysis of the entire sample.

We assume $H_{0}=70 \, \rm km \, s^{-1} \, Mpc^{-1}$,
$\Omega_{0}=0.3$, and $\Omega_{\Lambda} =0.7$.  The adopted initial
mass function (IMF) is that of \cite{kr01} in the mass range 0.1--100
$\textrm{M}_{\odot}$. 

\begin{table*}[!t]
\caption{Cluster properties \label{tab:clus}}
\centering
\begin{tabular}{lccccccccc} 
\bf{cluster} 	& \bf{RA} (J2000) & \bf{DEC} (J2000) & \bf{z} 	& \bf{HST imaging} 	& \bf{L$_{\rm X}$} (10$^{44}$erg s$^{-1}$)	& \bf{M$_{\rm 500}$}  (10$^{14}$M$_\odot$) & \bf{r$_{\rm 500}$}  (Mpc) & {\bf PA1} & {\bf PA2}    \\
\hline
MACS0717.5+3745   & 07:17:31.6  & +37:45:18 & 0.548 &    CLASH/HFF2   & 24.99$\pm$0.92& 24.9$\pm$2.7 & 1.69$\pm$0.06 & 020& 280 \\
MACS1423.8+2404   & 14:23:47.8  & +24:04:40 & 0.545 &    CLASH       & 13.96$\pm$0.52& 6.64$\pm$0.88   & 1.09$\pm$0.05 & 008& 088 \\
\hline
\end{tabular}
 \tablecomments{J2000 coordinates, redshift,  the main source of HST imaging, X-ray luminosity  \citep[from][]{mantz10}, M$_{500}$ (from M$_{\rm gas}$), r$_{500}$ and the two position angles. }
\end{table*}

\section{The Grism Lens-Amplified Survey from Space data set}
\label{sec:glass}

GLASS is a 140 orbit slitless spectroscopic survey with \HST{} in cycle 
21. It  has observed
the cores of 10 massive galaxy clusters with the  WFC3 NIR grisms G102 and G141 providing an uninterrupted 
wavelength coverage from 0.8$\mu$m to 1.7$\mu$m.  
Observations for GLASS were completed in January 2015.  Amongst the 10 GLASS clusters, 6 are targeted by the Hubble Frontier 
Fields (HFF; P.I. Lotz) and 8 by the Cluster Lensing And Supernova survey with Hubble (CLASH; P.I. Postman, \citealt{postman12}). Prior to each grism exposure, imaging through either F105W or F140W is obtained to assist 
the extraction of the spectra and the modeling of contamination from nearby objects on the sky.  The total exposure time per 
cluster is 10 orbits in G102 (with either F105W or F140W) and 4 orbits in G141 with F140W. Each cluster is observed at two position 
angles (PAs) approximately 90 degrees apart to facilitate clean extraction of the spectra for objects in crowded cluster fields.

\subsection{Data reduction}\label{sec:reduction}

The GLASS observations are designed after the 3D-HST observing strategy and were processed with an updated version of the 3D- HST reduction pipeline\footnote{\url{http://code.google.com/p/threedhst/}}  described by \cite{brammer12}. The updated pipeline combines the individual exposures into mosaics using AstroDrizzle \citep{gonzaga12}, replacing the MultiDrizzle package \citep{koekemoer03} used in earlier versions of the pipeline.

The direct images were sky subtracted by fitting
a 2nd order polynomial to each of the source-subtracted
exposures. Each exposure is then interlaced to a final
G102(G141) grism mosaic. Before sky-subtraction and interlacing, each individual exposure was checked and corrected for elevated backgrounds due to the He Earth-glow described by \cite{brammer14}. 
From the final mosaics, the spectra of  individual objects are extracted by predicting the position and extent of each two-dimensional spectrum based on the SExtractor \citep{bertin96} segmentation map combined with deep mosaic of the direct NIR GLASS and CLASH images. As this is done for each object, the contamination, i.e., the dispersed light from neighboring objects in the direct image field-of-view, can be estimated and accounted for. Full details on the sample selection, data observations and data reduction are given in \cite{treu15}, while a complete description of the 3D-HST image preparation pipeline, spectral extractions, and spectral fitting, is provided by Momcheva et al. (in prep.).

The spectra analyzed in this study were all visually inspected with the publicly available GLASS inspection GUI, GiG\footnote{\url{https://github.com/kasperschmidt/GLASSinspectionGUIs}} \citep{treu15},  
in order to  identify and flag erroneous models from the reduction, assess the degree of contamination in the spectra and flag and identify strong emission lines and the presence of a continuum. 

\subsection{Redshift determinations}
In order to determine redshifts, templates are compared to each of the four available grism spectra independently (G102 and G141 at two PAs each) to compute a posterior distribution function for the redshift. If available, photometric redshift distributions can be used as input priors to the grism fits in order to reduce computational time. Then, with the help of the publicly available GLASS inspection GUI for redshifts (GIGz, \citealt{treu15}
), we flag which grism fits are reliable or alternatively enter a redshift by hand if the redshift is misidentified by the automatic procedure.
Using GIGz we assign a quality flag to the redshift (4=secure; 3=probable; 2=possible; 1=tentative, but likely an artifact; 0=no-$z$). These quality criteria take into account the signal to noise ratio of the detection, the possibility that the line is a contaminant, and the identification of the feature with a specific emission line. This procedure is carried out independently by
at least two inspectors per cluster and then their outputs are combined
  \citep[see][for details]{treu15}.

We note that for \ma, one of the clusters analyzed in this paper (see \S\ref{sec:sample}), a redshift catalog was already published by \cite{ebeling14}. Considering only galaxies with quality flag $>$2.5, four objects are in common between the two catalogs (cross match within 1") and the reported redshift agree at the 10$^{-3}$ level, consistent with the resolution of the grism.

\subsection{The sample}
\label{sec:sample}

Even though all GLASS data have been  obtained and reduced, their
inspection and quality control is still underway, and is expected to
be completed  and released by Winter 2016 \citep{treu15}.  Among
the clusters for which quality control is sufficiently advanced for
this work, we select two with similar redshift,  \ma and \mb, whose properties are listed in  Table \ref{tab:clus}.

From the redshift catalogs, we extract  galaxies with secure
redshift (flag$\geq$2.5) and consider as cluster members galaxies with
redshift within $\pm$0.03 the cluster redshift.  Then, we select 
galaxies with visually detected \Ha in emission. Given the cluster
redshifts, \Ha is  found at an observed wavelength of
$\sim$10,100 \AA, and we therefore only exploit the G102 grism data in our analysis.

\begin{figure*}
\centering
\includegraphics[scale=0.4]{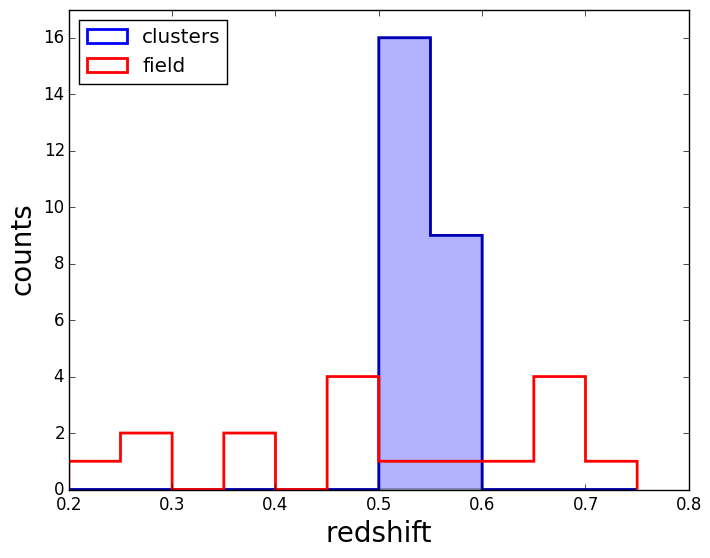}
\includegraphics[scale=0.4]{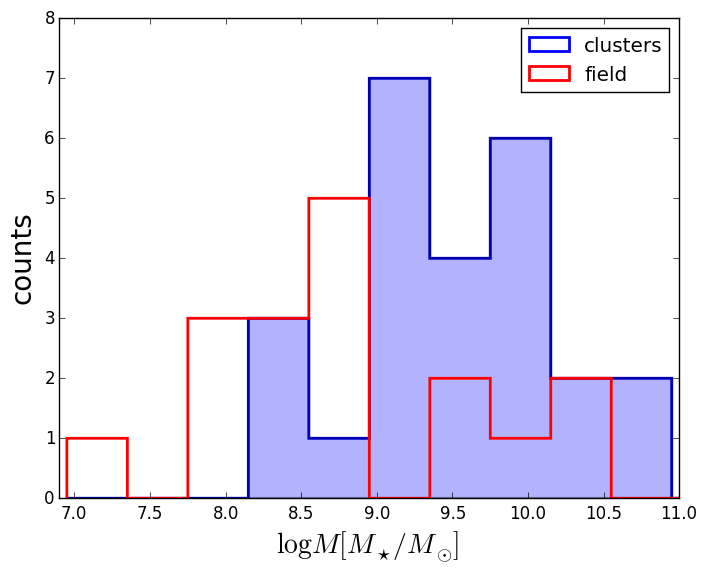}
\caption{Redshift (left) and mass (right) distributions for the cluster (blue, shade areas) and field (red) galaxies analyzed in this study. \label{fig:z_m_distr}}
\end{figure*}

We  assemble a control sample  which includes all galaxies with secure redshift, \Ha in emission detected in the G102 grism and redshift outside the cluster redshift intervals ($z<z_{cl}-0.03$ or $z>z_{cl}+0.03$). The field sample includes galaxies in the redshift range  0.2$<z<$0.7, and differences among these galaxies are therefore potentially  due to evolutionary effects, although the average redshift is very similar between the two samples. We note that we do not have additional information on the environments in which these galaxies reside, therefore they might be located in some groups.

Overall, our sample includes 15 cluster members and 9 field galaxies from the \ma field, and 10 cluster members and 8 field galaxies from the \mb field.

Stellar masses have been estimated using the broad-band CLASH photometry  \citep{postman12} and a set of templates, computed with standard spectral synthesis models \citep{bc03}, and fixing the redshift at the spectroscopic one.
As it was done in previous papers \citep[e.g.,][for details]{fontana04, fontana06, santini15}, we have used a range of exponential time scales ranging from $0.1$ to $\infty$~Gyr. A \cite{salpeter55} IMF, ranging over a set of metallicities (from Z = 0.02Z$_\odot$ to Z=2.5Z$_\odot$)  and dust extinction (0<E(B-V)<1.1, with a Calzetti extinction curve) has been initially chosen, and then converted to a  \cite{kr01} IMF.  We have also added emission lines in a self-consistent way, as  described by \cite{castellano14}, 
that provide an important contribution to our H$\alpha$-emitting galaxies. Uncertainties on the estimated masses have been derived by scanning (for each galaxy) all the templates and retaining only masses corresponding to models with $P(\chi^2)>0.1$ \citep{santini15}.

Figure \ref{fig:z_m_distr} shows the redshift and mass distribution for cluster and field samples separately.  We note that, while the mass range spanned in the two environments is similar, going from 10$^8$ to 10$^{11}$ M$_\odot$, 
 the field galaxies are systematically less massive than  cluster galaxies.  
 Therefore, in the following, when comparing cluster and field populations, differences might be due to the different mass distribution, and not only to purely environmental effects.

\section{Methodology}

\subsection{\Ha maps}

The combined spatial resolution on the WFC3 and of the grism yield a spectrum that can be seen as images of a galaxy taken at $\sim$24 \AA{} increments ($\sim$12 \AA{} after interlacing) and placed next to each other (offset by one pixel) on the detector. Thus, an emission feature in a high spatial resolution slitless spectrum is essentially an image of a galaxy at that wavelength. 

\begin{figure*}
\centering
\includegraphics[scale=0.45]{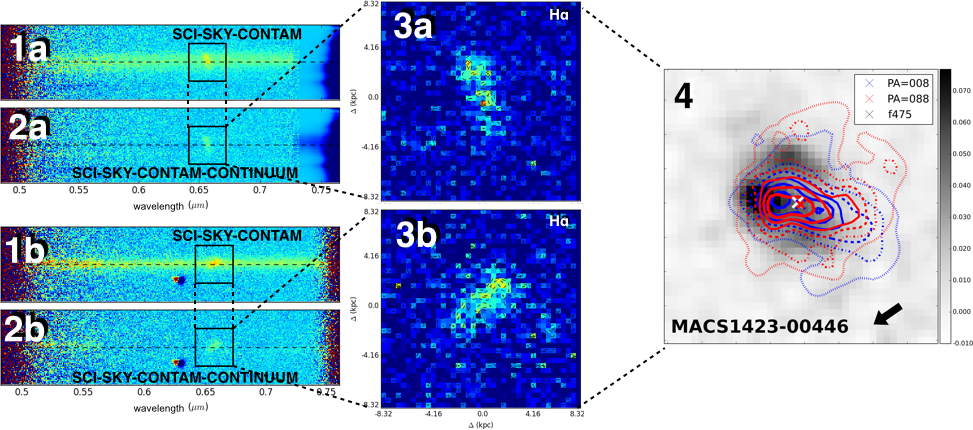}
\includegraphics[scale=0.45]{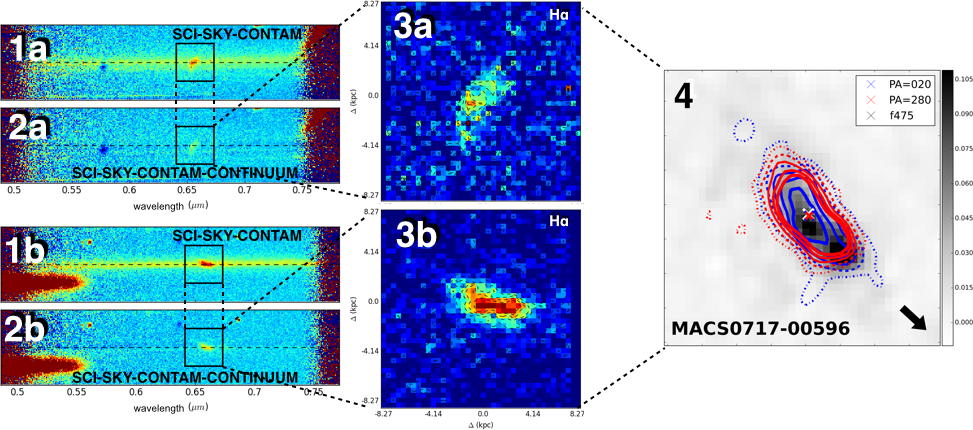}
\caption{Two examples of the procedure followed to derive $H_\alpha$ maps. Upper panels: MACS1423-00446, bottom panels: MACS0717-00596. $a$ and $b$ refer to the two distinct PAs of the same galaxy. Panels $1a$  and $1b$ show the rest-frame flux-calibrated galaxy 2D  spectra, after the sky background and the contamination have been subtracted. The dashed horizontal black lines show the y-position of the continuum. The box shows the position of the emission lines. Panels $2a$ and $2b$ show the  rest-frame 2D spectra after  the continuum has also been subtracted. Panels $3a$ and $3b$ show the map of the galaxy in the light of the $H_\alpha$ line. Panels $4$ show the two maps overplotted as contours on an image of the galaxy in the FW475 filter. Blue and red lines represent different levels of $\Sigma$SFR, as described in Fig.\ref{fig:c_size_Hec}. The arrows indicates the direction of the cluster center. See text for details. \label{fig:methods}}
\end{figure*}

Figure \ref{fig:methods} shows two examples of the procedure we
followed to create \Ha emission line maps and therefore SFR maps. In
the first steps, we treat the spectra coming from the different
exposures (one per PA) of each galaxy independently, and only in the
last step we combine them.  Both panels $1a$ and $1b$ show the
flux-calibrated galaxy 2D continuum spectra, after the sky background
and the contamination have been subtracted.  From two regions
contiguous to the \Ha emission we determine the $y$-position of the
peak of the continuum. This position will be needed to measure the
offset in the $y$-direction of the \Ha emission with respect to the
galaxy center in the light of the continuum. Subsequently, we subtract
the 2D stellar continuum model obtained by convolving the best-fit 1D
SED without emission lines with the actual 2D data, ensuring that all
model flux pixels are non-negative (panels $2a$ and $2b$). If the sky,
the contamination and the continuum were fit perfectly, we should be
left only with the flux coming from the emission lines. We find that
counts around the lines are slightly negative, suggesting that the
continuum subtraction is somewhat too aggressive.  Therefore, we
select a box just above and one just below the emission line and
measure the median flux which is further subtracted from the entire
spectrum. The residual is a map of the galaxy in the light of the \Ha
line (panels $3a$ and $3b$).  As a last step, we superimpose the \Ha
map onto an image of the galaxy taken with the F475W filter
(rest-frame UV).  We use F475W to map relatively recent ($\sim$100 Myr)  star formation,
as opposed to ongoing  ($\sim$10Myr) star formation traced by \Ha. 
To do so, we align each map to the image of the galaxy in the
light of the continuum, rotating each map by the angle of its PA,
keeping the $y$-offset unaltered with the respect to the continuum. On
the $x$-axis, there is a degeneracy between the spatial dimension and
the wavelength uncertainty, it is therefore not possible to determine very accurately
the central position of the \Ha map for each PA separately. Nonetheless, for the cases in
which spectra from both PAs are reliable (28/42), we use the fact that
the 2 PAs differ by almost $90\degree$, therefore the $x$-direction of
one spectrum roughly corresponds to the $y$-direction of the second
spectrum and vice-versa. We can therefore shift the two spectra
independently along their $x$-direction to force the center of the
emission of the two maps to coincide.  The results are shown in Figure
\ref{fig:methods}, panels $4$.
For the galaxies with reliable spectra in both PAs, we can also measure the real distance between the peak of the \Ha emission and the continuum emission, obtained as the quadratic sum of the two offsets. 

Finally, for  cluster galaxies, we also measure the magnitude of the offset between the \Ha and the continuum as projected along the cluster radial direction, determined by  the line connecting the cluster-center and the galaxy center in the continuum light. We assign a positive sign to the projected offset when the peak of the \Ha is between the cluster center and the peak of the continuum.  

\subsection{\Ha map sizes}

Since one of our aims is to compare the extent of \Ha light to the extent of the continuum light, we estimate galaxy sizes at  different wavelengths by measuring the second order moment of the light distribution, which gives us the width of the distribution and therefore the extension of the galaxy:
$$
\sigma=\sqrt{\frac{\sum_{i=1}^{N} \left[I(x_i) \cdot x_i^2\right]}{\sum_{i=1}^{N} I(x_i)}-\left[\frac{\sum_{i=1}^{N} \left[I(x_i) \cdot x_i\right]}{\sum_{i=1}^{N} I(x_i)}\right]^2}
$$
with $x_i$  along the spectrum, and $I(x_i)$  flux at the corresponding position. We measure sizes both along the $x-$ and $y-$direction. The average size is obtained by taking the mean of the two and summing errors in quadrature.   This adopted size definition is independent on the galaxy's center. 
When spectra from both  PAs are reliable, we take the average size and sum the errors in quadrature, after having checked that the measurements from the two PAs are consistent within the uncertainty.

 Besides on the \Ha light, we compute sizes both on the F475W filter, to map the star formation occurred in roughly  the last 100 Myr,  and on the F110W filter, which probes the rest-frame optical continuum and therefore maps the older stellar population.
We correct our size estimates for the point spread function (PSF) of our observations. We estimate the mean full width half maximum (FWHM) in each band by taking the average of the FWHM of 5 stars. We then subtract in quadrature the PSF (=FWHM/2.355) from the sizes. 
The values we obtain are $\sim$0.03$^{\prime\prime}$ in the F475W,  and $\sim$0.055$^{\prime\prime}$ in the F110W and G102 filters. We note that the PSF correction 
 is generally much smaller than the sizes we observe  therefore the impact of the correction is negligible.

We note that more robust measurements are currently underway for the entire GLASS sample and will be presented in a forthcoming paper.
\begin{table*}
\caption{Properties of galaxies in clusters
\label{tab:clu_gal}}
\centering
\setlength{\tabcolsep}{3pt}
\begin{tabular}{lllrrrrrrr}
\hline
\hline
  \multicolumn{1}{c}{obj\_name} &
  \multicolumn{1}{c}{RA} &
  \multicolumn{1}{c}{DEC} &
  \multicolumn{1}{c}{z} &
  \multicolumn{1}{c}{$\log$ M} &
  \multicolumn{1}{c}{EW} &
  \multicolumn{1}{c}{SFR} &
  \multicolumn{1}{c}{$\Sigma$SFR} &
  \multicolumn{1}{c}{dist$_{r_{500}}$} &
  \multicolumn{1}{c}{dist} \\
  \multicolumn{1}{c}{} &
  \multicolumn{1}{c}{(J2000)} &
  \multicolumn{1}{c}{(J2000)} &
  \multicolumn{1}{c}{} &
  \multicolumn{1}{c}{(M$_\odot$)} &
  \multicolumn{1}{c}{} &
  \multicolumn{1}{c}{(M$_\odot/yr$)} &
  \multicolumn{1}{c}{(M$_\odot/yr/kpc^2$)} &
  \multicolumn{1}{c}{} &
  \multicolumn{1}{c}{(kpc)} \\
\hline
  MACS0717-00173 & 07:17:35.64 & +37:45:59.2 & 0.556 & 9.5$^{+0.4}_{-0.5}$ 	& 23$\pm$2 	& 3.4$\pm$0.5 & 0.20$\pm$0.03 & 0.276 & 466\\
  MACS0717-00234 & 07:17:35.14 & +37:45:52.9 & 0.549 & 10.5$^{+0.1}_{-0.3}$ 	& 14.1$\pm$0.3 & 8.6$\pm$0.8 & 0.16$\pm$0.01 & 0.247 & 418\\
  MACS0717-00431 & 07:17:36.59 & +37:45:40.1 & 0.5495 & 8.9$^{+0.2}_{-0.3}$ 	& 29$\pm$3 	& 2.9$\pm$0.4 & 0.30$\pm$0.06 & 0.226 & 382\\
 MACS0717-00596 & 07:17:37.76 & +37:45:30.1 & 0.5475 & 9.9$^{+0.2}_{-0.4}$ 	& 65$\pm$2 	& 14.0$\pm$0.6 & 0.54$\pm$0.03 & 0.230 & 388\\
  MACS0717-00624 & 07:17:33.44 & +37:45:28.9 & 0.5725 & 9.0$^{+0.1}_{-0.3}$ 	& 35$\pm$4 	& 2.3$\pm$0.5 	& 0.15$\pm$0.04  &0.153 & 259\\
  MACS0717-00674 & 07:17:34.98 & +37:45:27.4 & 0.574 & 9.3$^{+0.2}_{-0.2}$ 	& 24$\pm$5	& 0.8$\pm$0.4 & 0.08$\pm$0.06 	& 0.152 & 257\\
  MACS0717-00977 & 07:17:38.86 & +37:45:20.0 & 0.567 & 9.3$^{+0.2}_{-0.3}$ 	& 18.5$\pm$0.6 & 4.8$\pm$0.7 & 0.10$\pm$0.02  & 0.248 & 419\\
  MACS0717-01208 & 07:17:32.79 & +37:44:41.3 & 0.5585 & 9.5$^{+0.3}_{-0.7}$ 	& 63$\pm$5 	& 5.1$\pm$0.4 & 0.53$\pm$0.06  & 0.062 & 105\\
  MACS0717-01305 & 07:17:35.55 & +37:44:41.8 & 0.5285 & 10.7$^{+0.2}_{-0.2}$ 	& 18.1$\pm$0.5 & 18$\pm$1 & 0.14$\pm$0.01 	 & 0.074 & 1256\\
  MACS0717-02181 & 07:17:31.51 & +37:44:13.1 & 0.564 & 9.0$^{+0.1}_{-0.4}$ 	& 11$\pm$4 	& 1.3$\pm$0.4 & 0.16$\pm$0.07  & 0.176 & 298\\
  MACS0717-02189 & 07:17:33.76 & +37:44:08.4 & 0.5275 & 10.0$^{+0.3}_{-0.4}$	 & 6.2$\pm$0.2 & 3.0$\pm$0.6 & 0.17$\pm$0.06	& 0.154 & 260\\
  MACS0717-02297 & 07:17:30.17 & +37:44:04.1 & 0.5485 & 9.2$^{+0.3}_{-0.3}$ 	& 19$\pm$1 	& 3.1$\pm$0.4 & 0.28$\pm$0.04 & 0.242 & 408\\
  MACS0717-02334 & 07:17:32.35 & +37:43:59.4 & 0.534 & 9.9$^{+0.2}_{-0.2}$ 	& 13.8$\pm$0.3 & 7.4$\pm$0.5 & 0.35$\pm$0.03 & 0.202 & 341\\
  MACS0717-02432 & 07:17:31.45 & +37:43:50.7 & 0.548 & 9.4$^{+0.4}_{-0.3}$ 	& 55$\pm$9 	& 2.9$\pm$0.5 & 0.17$\pm$0.03  & 0.249 & 420\\
  MACS0717-02574 & 07:17:31.76 & +37:43:33.8 & 0.5375 & 10.2$^{+0.6}_{-0.4}$ 	& 29$\pm$1 	& 9.1$\pm$0.7 & 0.25$\pm$0.02  & 0.302 & 511\\
  MACS1423-00152 & 14:23:49.65 & +24:05:43.0 & 0.53 & 9.0$^{+0.2}_{-0.3}$ 	& 21$\pm$2 	& 2.0$\pm$0.4 & 0.15$\pm$0.03 & 0.345 & 376\\
  MACS1423-00229 & 14:23:46.25 & +24:05:32.6 & 0.563 & 8.3$^{+0.1}_{-0.1}$ 	& 110$\pm$30 	& 4.1$\pm$0.5 & 0.28$\pm$0.05 & 0.313 & 342\\
  MACS1423-00310 & 14:23:45.62 & +24:05:27.3 & 0.53 & 9.9$^{+0.2}_{-0.4}$ 	& 26$\pm$1 	& 3.7$\pm$0.5 & 0.26$\pm$0.05  & 0.319 & 348\\
  MACS1423-00319 & 14:23:48.24 & +24:05:20.7 & 0.536 & 10.4$^{+0.2}_{-0.5}$ 	& 36$\pm$4 	& 3.0$\pm$0.7 & 0.12$\pm$0.03 & 0.198 & 215\\
  MACS1423-00446 & 14:23:45.18 & +24:05:16.4 & 0.548 & 9.9$^{+0.3}_{-0.4}$ 	& 57$\pm$2 	& 8.7$\pm$0.6 & 0.40$\pm$0.04  & 0.304 & 331\\
  MACS1423-00487 & 14:23:47.81 & +24:05:13.6 & 0.53 & 9.7$^{+0.3}_{-0.4}$ 	& 61$\pm$5 	& 7.1$\pm$0.6 & 0.38$\pm$0.03  & 0.161 & 175\\
  MACS1423-00831 & 14:23:49.24 & +24:04:52.3 & 0.575 & 8.5$^{+0.9}_{-0.5}$ 	& 41$\pm$4 	& 2.6$\pm$0.5 & 0.11$\pm$0.03 &0.081 & 89\\
  MACS1423-01516 & 14:23:48.56 & +24:04:14.6 & 0.54 & 10.6$^{+0.2}_{-0.3}$ 	& 27.1$\pm$0.3 & 14.8$\pm$0.7 & 0.48$\pm$0.03 & 0.190 & 208\\
  MACS1423-01253 & 14:23:53.13 & +24:04:29.4 & 0.556 & 9.1$^{+0.1}_{-0.2}$ 	& 13$\pm$1 	& 2.4$\pm$0.5 & 0.12$\pm$0.02 & 0.400 & 437\\
  MACS1423-01910 & 14:23:49.20 & +24:03:42.7 & 0.532 & 8.5$^{+0.2}_{-0.3}$ 	& 100$\pm$10 & 3.2$\pm$0.4 & 0.35$\pm$0.06 & 0.383 & 418\\
 \hline 
\end{tabular}
 \tablecomments{J2000 coordinates, redshift, stellar mass, \Ha equivalent width, SFR, $\Sigma$SFR, and distance from the cluster center (both in units of $r_{500}$ and in kpc). }
\end{table*}

\begin{table*}
\caption{Spatial properties of galaxies in clusters
\label{tab:clu_gal_2}}
\centering
\setlength{\tabcolsep}{3pt}
\begin{tabular}{lrrrrrrrrrrr}
\hline
\hline
  \multicolumn{1}{c}{obj\_name} &
  \multicolumn{1}{c}{offset 1} &
  \multicolumn{1}{c}{offset 2} &
  \multicolumn{1}{c}{r(\Ha)$_x$} &
  \multicolumn{1}{c}{r(\Ha)$_y$} &
  \multicolumn{1}{c}{r(F110W)$_x$} &
  \multicolumn{1}{c}{r(F110W)$_y$} &
  \multicolumn{1}{c}{r(F475W)$_x$} &
  \multicolumn{1}{c}{r(F475W)$_y$} &
  \multicolumn{1}{c}{A} &
  \multicolumn{1}{c}{B} &
  \multicolumn{1}{c}{$\theta$} \\
  \multicolumn{1}{c}{} &
  \multicolumn{1}{c}{(kpc)} &
  \multicolumn{1}{c}{(kpc)} &
  \multicolumn{1}{c}{(kpc)} &
  \multicolumn{1}{c}{(kpc)} &
  \multicolumn{1}{c}{(kpc)} &
  \multicolumn{1}{c}{(kpc)} &
  \multicolumn{1}{c}{(kpc)} &
  \multicolumn{1}{c}{(kpc)} &
  \multicolumn{1}{c}{(kpc)} &
  \multicolumn{1}{c}{(kpc)} &
  \multicolumn{1}{c}{(deg)} \\
  \hline
  MACS0717-00173 & - 				& 0.4$\pm$0.1 		& 4.1$\pm$0.3 		& 1.0$\pm$0.2 	&1.63  &0.82 &3.77&1.24 & 3.25 & 1.62 & -47.6 \\
  MACS0717-00234 & - 				& -0.4$\pm$0.1 	& 3.7$\pm$0.5 		& 2.0$\pm$0.8 	& 3.37&2.44 & 3.17 &2.13 & 5.32 & 3.2 & -16.3 \\
  MACS0717-00431 & -0.43$\pm$0.1 	& 0.18$\pm$0.07 	& 1.38$\pm$0.09 	& 1.0$\pm$0.2 	& 1.09& 0.98 &1.62 & 1.45 & 1.81 & 1.61 & 17.4 \\
 MACS0717-00596 &  0.11$\pm$0.05 	& -0.16$\pm$0.03 	& 3.2$\pm$0.4 		& 2.6$\pm$0.3 	& 2.84&1.82& 3.39 & 2.22 & 3.5 & 2.39 & -61.8 \\
  MACS0717-00624 &  -0.4$\pm$0.2 	& -0.5$\pm$0.3 	& 3.7$\pm$0.8 		& 4$\pm$1 	& 1.88&1.20&0.80 & 0.54 & 2.58 & 1.83 & -81.5 \\
  MACS0717-00674 & 0.2$\pm$0.4 		& 0.7$\pm$0.3 		& 4.3$\pm$0.3 		& 3$\pm$1 	&2.28 &2.11&5.69 & 6.28 & 1.96 & 1.51 & -58.4\\
  MACS0717-00977 &   -				& 0.1$\pm$0.2 		& 3.3$\pm$0.7 		& 3.7$\pm$0.6 	& 3.06&2.71&3.67 & 3.45 & 4.28 & 3.55 & -80.3 \\
  MACS0717-01208 & -0.19$\pm$0.07 	& 0.31$\pm$0.04 	& 3.9$\pm$0.6 		& 3.2$\pm$0.5 	&3.20&3.11& 1.64 & 1.95 & 1.95 & 1.57 & -44.7 \\
  MACS0717-01305 &  -1.6$\pm$0.1 	& -1.0$\pm$0.2 	& 3.1$\pm$0.3 		& 2.9$\pm$0.4 	&3.71 &3.73&3.57 & 3.12 & 7.06 & 5.4 & 23.5 \\
  MACS0717-02181 & -0.2$\pm$0.2 		& 0.1$\pm$0.2 		& 4$\pm$1 		& 3.7$\pm$0.9 	&1.07&0.91& - & 4.61 & 1.68 & 1.52 & -47.0 \\
  MACS0717-02189 &  -1.9$\pm$0.3 	& -0.1$\pm$0.1 	& 3.2$\pm$0.3 		& 2.7$\pm$0.4 	&2.98&3.06& 3.05 & 1.40 & 3.5 & 1.5 & -55.1 \\
  MACS0717-02297 & -0.06$\pm$0.09 	&  - 				& 2$\pm$1 		& 1.5$\pm$0.6 	& 0.94&0.97&2.89 & 1.40 & 2.01 & 1.7 & -14.1 \\
  MACS0717-02334 &  - 				& 0.62$\pm$0.05 	& 1$\pm$1.0 		& 1.7$\pm$0.6 	&2.77&2.14& 2.23 & 1.52 & 3.35 & 2.02 & -74.7 \\
  MACS0717-02432 & 0.4$\pm$0.2 		& - 				& 1$\pm$1 		& 0.8$\pm$0.1 	&2.68&0.59& 2.89 & 0.73 & 3.69 & 1.48 & -80.5 \\
  MACS0717-02574 &  0.11$\pm$0.09 	&  - 				& 3.6$\pm$0.3 		& 3.5$\pm$0.3 	&-&-& 3.93 & 1.25 & 7.34 & 1.57 & -74.0 \\
  MACS1423-00152 &  -0.2$\pm$0.2 	& - 				& 6$\pm$2 		& 5$\pm$2 	&1.16&1.12& 0.98 & 1.03 & 0.71 & 0.61 & 3.1 \\
  MACS1423-00229 &  0.23$\pm$0.08 	& -0.1$\pm$0.1 	& 4.4$\pm$0.4 		& 3$\pm$1 	&3.09&1.68& 3.01 & 0.61 & 1.28 & 1.05 & -6.0 \\
  MACS1423-00310 &  0.16$\pm$0.07 	& 0.5$\pm$0.2 		& 4.2$\pm$0.5 		& 2.8$\pm$0.7 &2.02&2.10& 1.85 & 2.49 & 2.17 & 1.96 & 16.7 \\
  MACS1423-00319 &  - 				& 0.11$\pm$0.09 	& 4.5$\pm$0.5 		& 5.0$\pm$0.5 	&2.92&2.14& 1.11 & 0.84 & 3.65 & 3.18 & 4.0 \\
  MACS1423-00446 &  -0.28$\pm$0.05 	& -0.18$\pm$0.06 	& 3.2$\pm$0.3 		& 3.3$\pm$0.3 	&2.60&1.95& 2.92 & 2.09 & 6.71 & 2.56 & 86.4 \\
  MACS1423-00487 &  - 				& 0.42$\pm$0.07 	& 3.1$\pm$0.3 		& 1.4$\pm$0.1 	&3.03&1.72& 1.90 & 0.89 & 0.81 & 0.44 & -55.4 \\
  MACS1423-00831 &  -0.1$\pm$0.2 	& -0.2$\pm$0.3 	& 4.1$\pm$0.6 		& 4.9$\pm$0.5 	&3.06&1.75& 1.88 & 1.13 & 1.0 & 0.8 & -12.0 \\
  MACS1423-01253 &  -0.0$\pm$0.2 	& - 				& 4.0$\pm$0.8 		& 3.2$\pm$0.9 	&2.09&1.75& 1.25 & 1.09 & 2.59 & 1.73 & 70.7 \\
  MACS1423-01516 &  -0.29$\pm$0.04 	& -0.76$\pm$0.06 	& 3.8$\pm$0.2 		& 3.7$\pm$0.2 	&2.50&2.33& 2.83 & 2.90 & 1.57 & 1.37 & 75.6 \\
  MACS1423-01910 & -0.15$\pm$0.05 	& 0.5$\pm$0.1 		& 2$\pm$1 		& 4$\pm$1 	& 1.62&1.16&1.45 & 1.90 & 0.68 & 0.53 & 22.1\\
 \hline 
\end{tabular}
 \tablecomments{Offsets between the \Ha emission and the continuum emission along the two directions, PSF-corrected sizes of the \Ha maps, sizes of the rest-frame optical and UV-continuum (as measured on the F110W and F475W filter), Kron sizes as measured by Sextractor, and galaxy inclination.  Reported offsets are along the y-direction of the corresponding PA. The orientation of the offset  (counterclockwise from North) is $\theta=-PA1-44.69$. Sizes are given both along the $x-$ and $y$-direction. The average size can be obtained by taking the mean of the two. Errors on the F110W,  F475W and Kron sizes are very small and dominated by systematics, therefore we do not report them. }
\end{table*}

\begin{table*}
\caption{Properties of galaxies in the field
\label{tab:fie_gal}}
\centering
\setlength{\tabcolsep}{3pt}
\begin{tabular}{lllrrrrr}
\hline
\hline
  \multicolumn{1}{c}{obj\_name} &
  \multicolumn{1}{c}{RA} &
  \multicolumn{1}{c}{DEC} &
  \multicolumn{1}{c}{z} &
  \multicolumn{1}{c}{$\log$ M} &
  \multicolumn{1}{c}{EW} &
  \multicolumn{1}{c}{SFR} &
  \multicolumn{1}{c}{$\Sigma$SFR} \\
  \multicolumn{1}{c}{} &
  \multicolumn{1}{c}{(J2000)} &
  \multicolumn{1}{c}{(J2000)} &
  \multicolumn{1}{c}{} &
  \multicolumn{1}{c}{(M$_\odot$)} &
  \multicolumn{1}{c}{} &
  \multicolumn{1}{c}{(M$_\odot/yr$)} &
  \multicolumn{1}{c}{(M$_\odot/yr/kpc^2$)} \\
\hline
  MACS0717-00236 & 07:17:34.48 & +37:45:52.1 & 0.39 & 10.2$^{+0.3}_{-0.3}$ & 11.3$\pm$0.2 & 15$\pm$1 & 0.29$\pm$0.03   \\
  MACS0717-00450 & 07:17:37.39 & +37:45:34.9 & 0.5965 & 8.5$^{+0.2}_{-0.5}$ & 90$\pm$40 & 2.4$\pm$0.5 & 0.13$\pm$0.04\\
  MACS0717-01234 & 07:17:37.56 & +37:45:09.3 & 0.2295 & 8.1$^{+0.3}_{-0.3}$ & 110$\pm$10 & 7.2$\pm$1.6 & 0.6$\pm$0.2  \\
  MACS0717-01416 & 07:17:39.72 & +37:44:50.7 & 0.5095 & 9.7$^{+0.3}_{-0.3}$ & 32$\pm$1 & 6.0$\pm$0.4 & 0.51$\pm$0.03  \\
  MACS0717-01477 & 07:17:29.74 & +37:44:46.2 & 0.45 & 9.5$^{+0.1}_{-0.3}$ & 9.6$\pm$0.6 & 9$\pm$1 & 0.13$\pm$0.02   \\
  MACS0717-01589 & 07:17:32.33 & +37:44:37.6 & 0.385 & 8.0$^{+0.9}_{-0.1}$ & 40$\pm$2 & 10$\pm$1 & 0.24$\pm$0.02  \\
  MACS0717-02371 & 07:17:31.95 & +37:44:01.7 & 0.263 & 8.8$^{+0.3}_{-0.5}$ & 17$\pm$3 & 2.7$\pm$0.8 & 0.4$\pm$0.2  \\
  MACS0717-02390 & 07:17:34.63 & +37:43:53.8 & 0.473 & 8.7$^{+0.1}_{-0.2}$ & 35$\pm$2 & 6$\pm$1 & 0.17$\pm$0.05   \\
  MACS0717-02445 & 07:17:32.38 & +37:43:51.0 & 0.49 & 8.7$^{+0.1}_{-0.2}$ & 90$\pm$30 & 2.8$\pm$0.5 & 0.20$\pm$0.06   \\
  MACS1423-00246 & 14:23:49.68 & +24:05:33.6 & 0.66 & 7.0$^{+0.9}_{-0.7}$ & 70$\pm$50 & 0.4$\pm$0.2 & 0.12$\pm$0.06  \\
  MACS1423-00256 & 14:23:45.35 & +24:05:31.4 & 0.71 & 7.8$^{+0.8}_{-0.5}$ & -  & 0.2$\pm$0.2 & 0.1$\pm$0.1  \\
  MACS1423-00463 & 14:23:46.44 & +24:05:15.4 & 0.62 & 8.5$^{+0.3}_{-0.1}$ & 77$\pm$9 & 1.8$\pm$0.4 & 0.13$\pm$0.03   \\
  MACS1423-00610 & 14:23:49.10 & +24:05:02.6 & 0.655 & 10.3$^{+0.2}_{-0.3}$ & 24.3$\pm$0.5 & 9.0$\pm$0.7 & 0.22$\pm$0.02 \\
  MACS1423-00677 & 14:23:45.68 & +24:04:55.2 & 0.665 & 8.9$^{+0.2}_{-0.2}$ & 49$\pm$5 & 4.3$\pm$0.4 & 0.33$\pm$0.05  \\
  MACS1423-01729 & 14:23:46.13 & +24:04:00.2 & 0.455 & 8.7$^{+0.4}_{-0.4}$ & 42$\pm$4 & 3.4$\pm$0.5 & 0.28$\pm$0.06  \\
  MACS1423-01771 & 14:23:44.43 & +24:03:56.1 & 0.65 & 10.1$^{+0.2}_{-0.2}$ & 38$\pm$1 & 11.8$\pm$0.7 & 0.34$\pm$0.02  \\
  MACS1423-01972 & 14:23:48.07 & +24:03:34.6 & 0.278 & 8.3$^{+0.5}_{-0.3}$ & 140$\pm$50 & 3.4$\pm$0.6 & 0.9$\pm$0.2   \\
\hline
\end{tabular}
\tablecomments{J2000 coordinates, redshift, stellar mass, \Ha equivalent width, SFR, and $\Sigma$SFR. }
\end{table*}

\begin{table*}
\caption{Spatial properties of galaxies in the field
\label{tab:fie_gal_2}}
\centering
\setlength{\tabcolsep}{3pt}
\begin{tabular}{lrrrrrrrrrrrrrrrrr}
\hline
\hline
  \multicolumn{1}{c}{obj\_name} &
  \multicolumn{1}{c}{offset 1} &
  \multicolumn{1}{c}{offset 2} &
  \multicolumn{1}{c}{r(\Ha)$_x$} &
  \multicolumn{1}{c}{r(\Ha)$_y$} &
  \multicolumn{1}{c}{r(F110W)$_x$} &
  \multicolumn{1}{c}{r(F110W)$_y$} &
   \multicolumn{1}{c}{r(F475W)$_x$} &
  \multicolumn{1}{c}{r(F475W)$_y$} &
  \multicolumn{1}{c}{A} &
  \multicolumn{1}{c}{B} &
  \multicolumn{1}{c}{$\theta$} \\
   \multicolumn{1}{c}{} &
  \multicolumn{1}{c}{(kpc)} &
  \multicolumn{1}{c}{(kpc)} &
  \multicolumn{1}{c}{(kpc)} &
  \multicolumn{1}{c}{(kpc)} &
  \multicolumn{1}{c}{(kpc)} &
  \multicolumn{1}{c}{(kpc)} &
  \multicolumn{1}{c}{(kpc)} &
  \multicolumn{1}{c}{(kpc)} &
  \multicolumn{1}{c}{(kpc)} &
  \multicolumn{1}{c}{(kpc)} &
  \multicolumn{1}{c}{(deg)} \\
\hline
  MACS0717-00236  &0.1$\pm$0.1 & 0.6$\pm$0.1 & 3.8$\pm$0.1 & 3.3$\pm$0.2 	&3.09&2.62& 3.39 & 2.85 & 4.85 & 3.59 & 47.9  \\
  MACS0717-00450  &-2.2$\pm$0.5 & 0.8$\pm$0.2 & 3.2$\pm$0.5 & 0.8$\pm$0.2 	&3.76&0.83& 3.72 & 3.82 & 3.14 & 1.85 & 4.8  \\
  MACS0717-01234  &0$\pm$1 & 0.3$\pm$0.2 & 2$\pm$2 & 0.7$\pm$0.2 		&1.27&1.13& 1.12 & 1.01 & 2.88 & 1.31 & 76.2 \\
  MACS0717-01416  &- & -0.24$\pm$0.05 & 5$\pm$1 & 5$\pm$1				&-&- & 2.25 & 2.56 & 1.94 & 1.82 & -0.9 \\
  MACS0717-01477 &-1.3$\pm$0.3 & -0.2$\pm$0.2 & 3.5$\pm$0.4 & 3.7$\pm$0.3	&3.29&2.90 & 3.50 & 3.29 & 5.29 & 4.08 & -19.3  \\
  MACS0717-01589 & - & -0.1$\pm$0.1 & 2.4$\pm$0.7 & 1$\pm$1 				&3.29&3.37& 3.40 & 3.59 & 3.91 & 3.5 & -29.1 \\
  MACS0717-02371 & 0.4$\pm$0.4 & -0.1$\pm$0.7 & 1.4$\pm$0.6 & 1.5$\pm$0.9 	& 0.63&1.04&0.69 & 0.89 & 1.34 & 1.15 & -27.2  \\
  MACS0717-02390 & 0.1$\pm$0.1 & 0.1$\pm$0.1 & 1.5$\pm$0.8 & 0.9$\pm$0.1 	&2.03&1.63& 3.90 & 2.72 & 6.32 & 3.36 & -86.5  \\
  MACS0717-02445 &  0.1$\pm$0.1 & -0.1$\pm$0.3 & 4$\pm$1 & 3$\pm$1 		&1.61&0.48& 2.88 & 0.96 & 2.43 & 1.25 & -77.7  \\
  MACS1423-00246 &  -0.2$\pm$0.2 &  - & - & 5$\pm$1						&-&-& 5.67 & 7.03 & 0.7 & 0.44 & -77.6 \\
  MACS1423-00256 &   - & -1.2$\pm$0.7 & - & 0.9$\pm$0.3	 				&1.41&4.48& 4.35 & 2.51 & 2.63 & 2.27 & -11.5 \\
  MACS1423-00463 &  0.3$\pm$0.2 & -  & 34$\pm$1 & 3$\pm$2 				&2.43&2.36& 2.14 & 1.46 & 1.58 & 0.86 & 32.7  \\
  MACS1423-00610 &  -1.2$\pm$0.1 & 0.11$\pm$0.09 & 4.0$\pm$0.3 & 3.2$\pm$0.3 &3.59&2.53& 3.79 & 2.32 & 1.72 & 0.71 & 74.3  \\
  MACS1423-00677 & 0.01$\pm$0.05 & -0.3$\pm$0.1 & 4.0$\pm$0.3 & 3.6$\pm$0.4 &2.17 &1.73&2.16 & 3.52 & 1.34 &1.17 & 20.5 \\
  MACS1423-01729 & -0.1$\pm$0.1 & -0.7$\pm$0.2 & 3$\pm$1 & 3.4$\pm$0.6 	&1.86&1.65& 1.88 & 1.81 & 1.84 & 1.43 & 47.5  \\
  MACS1423-01771 &  - & 0.07$\pm$0.07 & 3.7$\pm$0.3 & 3.4$\pm$0.3 			&3.12&2.84& 3.35 & 3.26 & 4.22 & 2.49 & 34.3  \\
  MACS1423-01972 &  0.2$\pm$0.1 & -0.1$\pm$0.1 & 2.3$\pm$0.8 & 0.9$\pm$0.3 	&0.78&1.19& 1.21 & 1.25 & 4.05 & 1.67 & -40.8  \\
\hline
\end{tabular}
 \tablecomments{Offsets between the \Ha emission and the continuum emission along the two directions, PSF-corrected sizes of the \Ha maps, sizes of the rest-frame optical and UV-continuum (as measured on the F110W and F475W filter), Kron sizes as measured by Sextractor, and galaxy inclination. 
 Reported offsets are along the y-direction of the corresponding PA. The orientation of the offset  (counterclockwise from North) is $\theta=-PA1-44.69$.  Sizes are given both along the $x-$ and $y$-direction. The average size can be obtained by taking the mean of the two.  Errors on the F110W,  F475W and Kron sizes are very small and dominated by systematics, therefore we do not report them. }
\end{table*}

\subsection{SFRs and EW(\Ha)s}

From the \Ha maps we also derive SFRs.
We use the conversion factor derived by \cite{kennicutt94}  and \cite{madau98}: 
$$\textrm{SFR} [\textrm{M}_\odot \, \textrm{yr}^{-1}] = 5.5 \times 10^{-42}\textrm{L}(H\alpha)[\textrm{erg}\,  \textrm{s}^{-1}]
$$
valid for a \cite{kr01} IMF.  
We compute both the surface  SFR density ($\Sigma$\textrm{SFR}, $\textrm{M}_\odot \, \textrm{yr}^{-1}\, \textrm{kpc}^{-2}$) and the total SFRs ($\textrm{M}_\odot \, \textrm{yr}^{-1}$), separately for the spectra coming from the two PAs and then we combine them taking the mean values. Errors are summed in quadrature.  The total SFRs are obtained summing the surface  SFR density  within the Kron radius\footnote{Kron radii are measured by Sextractor from a combined NIR image.}  of the galaxy. 

There are two major limitations when using \Ha as SFR estimator: the contamination by the [NII] line doublet, and uncertainties in the extinction corrections to be applied to each galaxy.

To correct for the scatter due to the [NII] contamination,  we apply the locally calibrated correction factor given by \cite{james05}. As opposed to  previous works which considered only central regions, these authors developed a method which takes into account the variation of the \Ha/[NII] with radial distance from the galaxy center, finding an average value of \Ha/(\Ha + [NII])= 0.823. This approach is appropriate given our goal to investigate extended emission.

\begin{figure}
\centering
\includegraphics[scale=0.45]{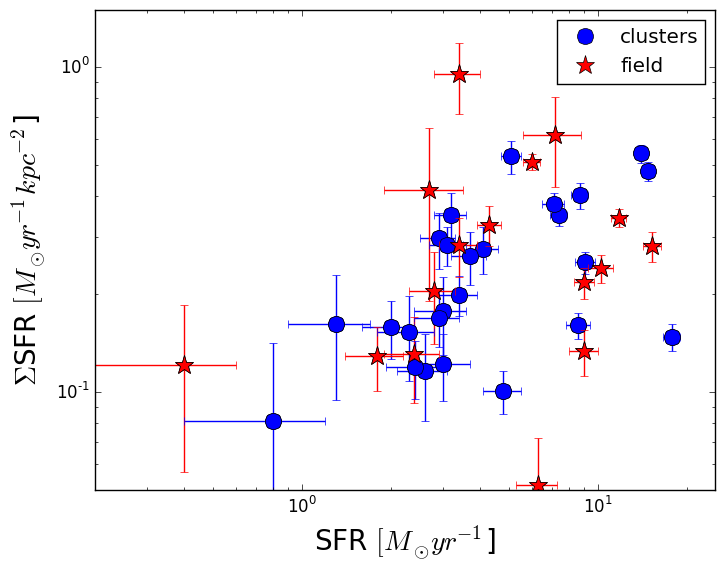}
\caption{$\Sigma$SFR-SFR for cluster (blue) and field (red) galaxies in our sample.
\label{fig:SFR_SFRd}}
\end{figure}

The second major problem when deriving SFR(\Ha) is the effect of dust extinction. Star formation normally takes place in dense and dusty molecular clouds, so a significant fraction of the emitted light from young stars is absorbed by the dust and re-emitted at rest-frame IR wavelengths. \cite{hopkins01} modeled a SFR-dependent attenuation by dust, characterized by the Calzetti reddening curve of the form
\begin{eqnarray*}
\log(\textrm{SFR}_i) =   \log(\textrm{SFR}_o(H\alpha)) + 2.614 \times \\
  \log \left[ \frac{0.797\times \log(\textrm{SFR}_i ) + 3.834}{2.88} \right]
\end{eqnarray*}
which allows us to estimate attenuation and intrinsic SFR, even for  observations of a single \Ha emission line. 
We use this correction to obtain the intrinsic SFRs. 

Figure \ref{fig:SFR_SFRd} correlates the total $\Sigma$SFRs to the  SFR and shows  that our   $\Sigma$SFR limit is around $10^{-1}$ M$_\odot$ yr$^{-1}$ kpc$^{-2}$ for SFR$\sim1 M_\odot yr^{-1}$.

Finally, we also compute \Ha equivalent widths (EW(\Ha)) from the collapsed 2D spectra.
We define the line profile by adopting a fixed rest frame wavelength range, centered on the theoretical wavelength, 6480-6650 \AA{}, and then  obtain the line flux, $f_{\textrm{line}}$, by summing the flux within the line. 
The continuum is defined by two regions of 100 \AA{} located at the two extremes of the line profile. We fit a straight line to the average continuum in the two regions and  sum the flux below the line, to obtain $f_{\textrm{cont}}$. The rest-frame  EW(\Ha) is therefore defined by
$$
\textrm{EW}(H_\alpha) =  \frac{f_{\textrm{line}}}{f_{\textrm{cont}}\times(1+z)}
$$
We note that our approach ignores underlying \Ha absorption. As usual, when two spectra for the same galaxy are reliable, the final value is given by the average of the two EW estimates, and the error is obtained by summing in quadrature the individual errors. The measurements from the two PAs are consistent within the uncertainty. Otherwise we just use a single spectrum.

\section{Results}\label{sec:results}

\begin{figure*}
\centering
\includegraphics[scale=0.65]{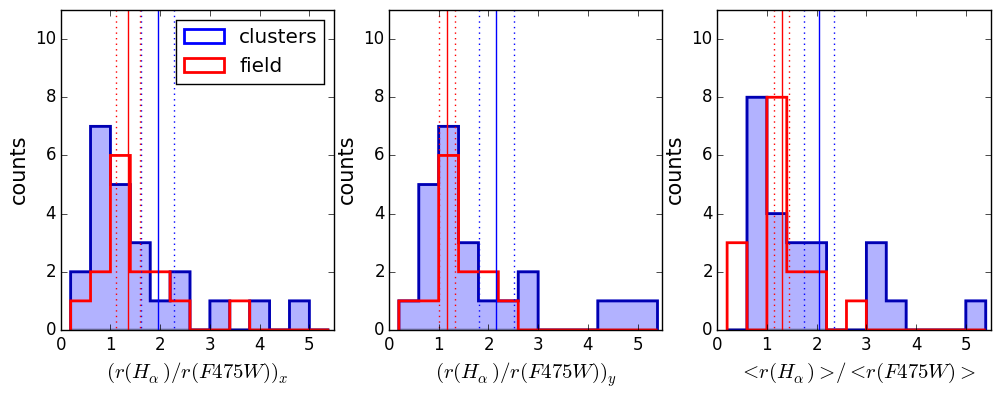}
\includegraphics[scale=0.65]{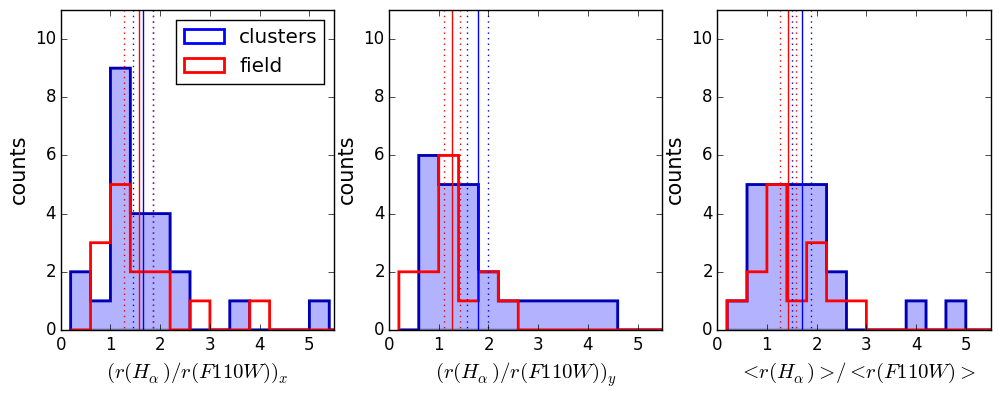}
\caption{Distribution of the ratio of the size of the \Ha emission to the size as measured from the F475W filter (upper panel) and from the F110W filter (bottom panel). Left and central panels show the sizes in the two directions separately, right panel shows the mean size. Vertical solid and dotted lines represent the means with errors. Both in clusters and in the field mean values are larger than 1, suggesting the inside-out growth, but when using the F475W, tracer of the  recent star formation, in clusters mean values are even larger, pointing at cluster specific processes at work. 
\label{fig:size}}
\end{figure*}

Tables \ref{tab:clu_gal} to \ref{tab:fie_gal_2} summarize the properties of the galaxies in our cluster and field samples, respectively. They include  galaxy positions, redshifts, stellar masses,  EW(\Ha)s, SFRs, $\Sigma$SFRs, sizes in different bands (F475W, F110W and \Ha) along both the $x-$ and $y-$ direction,  the offset between the peak of the light in \Ha and in the  rest-frame UV continuum, and, for clusters, the cluster-centric distances (both in kpc and in units of $r_{500}$).

Figure \ref{fig:size} shows the distribution of the ratio of the size as measured from the \Ha light (r(\Ha)) to the size as measured from  the rest-frame UV continuum (r(F745W)) and rest-frame optical continuum (r(F110W)), both for the two directions separately and for the mean sizes, obtained as average between the two directions.  
 We therefore compare the currently star forming regions to the younger stellar population (traced by the observed F475W continuum) and to the older one (traced by the observed F110W continuum). In a forthcoming analysis we will also compare  \Ha maps to maps of the even older stellar populations, as traced by the rest-frame Infrared. 

In both environments, there is no  preferential axis for the \Ha emission.  Ratios obtained using the F110W and the F475W filters agrees within the errors,  in both clusters and field. Looking at the continuum sizes, we do not detect strong trends with the wavelengths, most likely   because of the our currently small number statistics.  Distributions peak around 1, showing that many galaxies have comparable sizes in the line and continuum. However, distributions are slightly skewed toward values $>1$.   In cluster galaxies mean size ratios are systematically slightly larger than field galaxies when the F475W is used, but not when the F110W filter is considered  (clusters: $(r(H\alpha)/r(F475W))_x=1.8\pm0.3$, $(r(H\alpha)/r(F475W))_y=2.2\pm0.3$, $\langle r(H\alpha)\rangle/ \langle r(F745W)\rangle=1.9\pm0.3$; $(r(H\alpha)/r(F110W))_x=1.6\pm0.2$, $(r(H\alpha)/r(F110W))_y=1.8\pm0.2$, $\langle r(H\alpha)\rangle/ \langle r(F1110W)\rangle=1.7\pm0.2$;  field: ($r(H\alpha)/r(F475W))_x=1.3\pm0.2$, $(r(H\alpha)/r(F475W))_y=1.2\pm0.2$, $\langle r(H\alpha)\rangle/ \langle r(F745W)\rangle=1.3\pm0.1$; $(r(H\alpha)/r(F110W))_x=1.6\pm0.3$, $(r(H\alpha)/r(F110W))_y=1.6\pm0.2$, $\langle r(H\alpha)\rangle/ \langle r(F110W)\rangle=1.4\pm0.1$). Mean values for cluster galaxies are driven by a subpopulation of galaxies ($\sim$ 20\%) which present \Ha emission  at least two-three time as extended as the light in the rest-frame UV continuum. No such examples are present in the field.  This  might suggest that in all environments star formation is probably occurring over a larger area than that of the  recent star formation ($\sim100$ Myr), but in clusters there might be some additional mechanisms that are stripping the gas and star formation 
is continuing in the stripped material. 
A Kolomgorov-Smirnov (K-S) test can not reject the hypothesis that the distributions of the two samples are the same, giving probabilities lower than 80\% in all the three cases. We note that both samples are quite small, which might explain why the K-S test is inconclusive.

We use the information in the top right panel of Fig. \ref{fig:size} to group galaxies into different classes, as described in the next subsection.

\subsection{Maps of \Ha and continuum emission}
\begin{figure*}
\centering
  \includegraphics[scale=0.19]{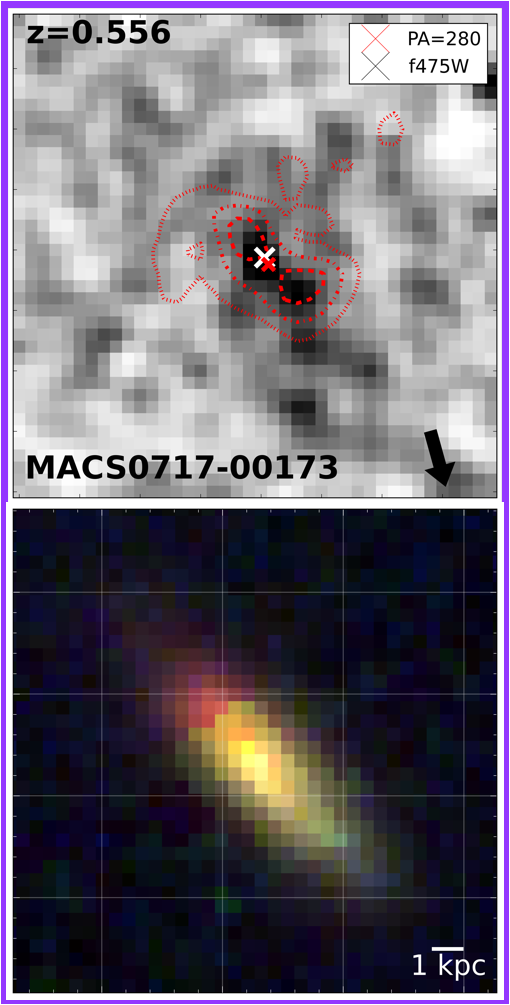}
  \includegraphics[scale=0.19]{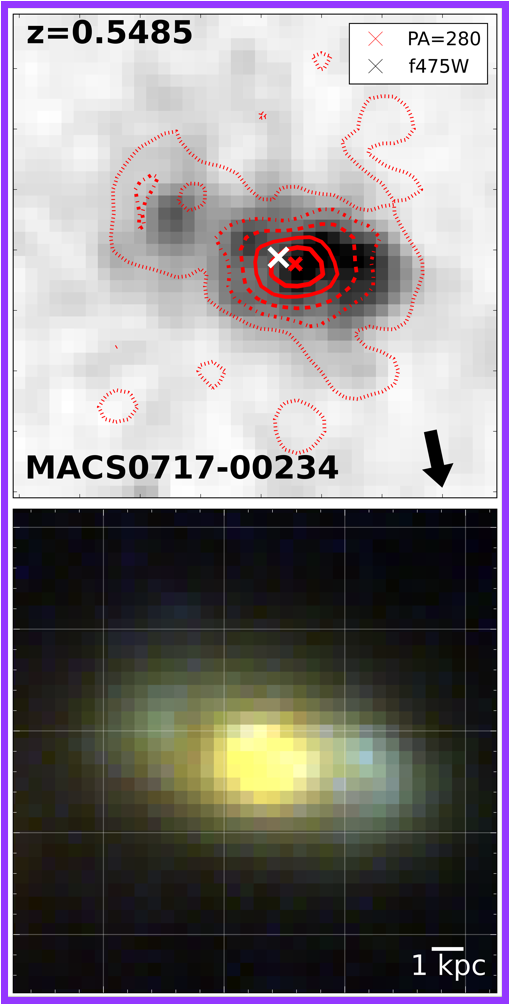}
  \includegraphics[scale=0.19]{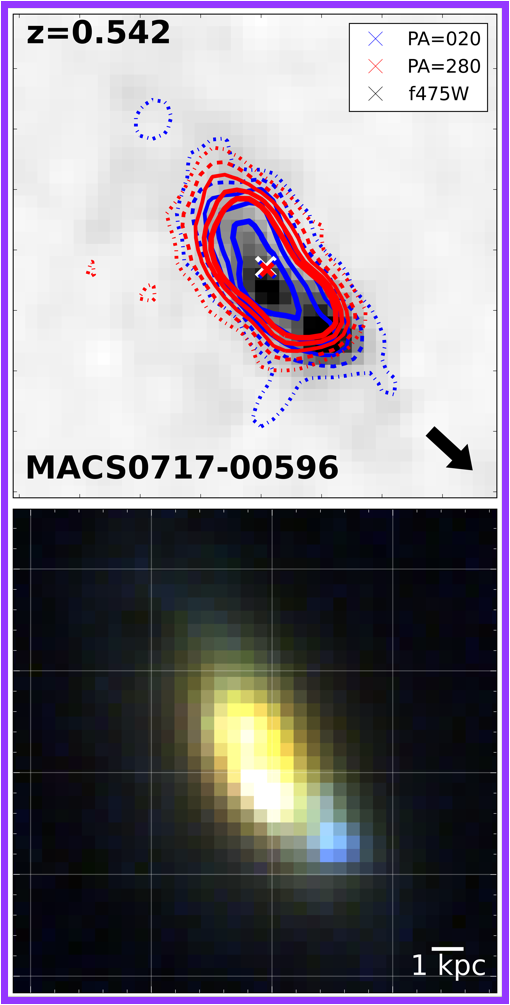}
  \includegraphics[scale=0.19]{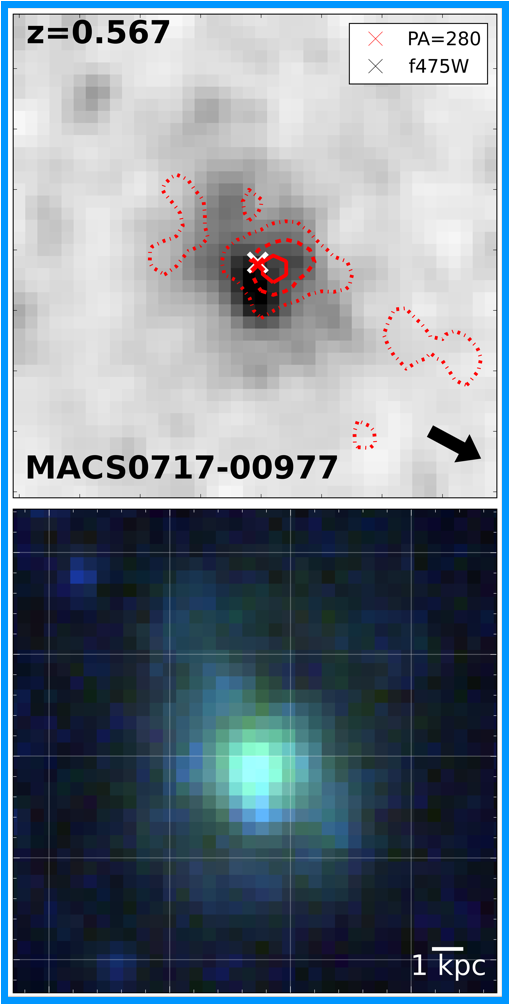}
  \includegraphics[scale=0.19]{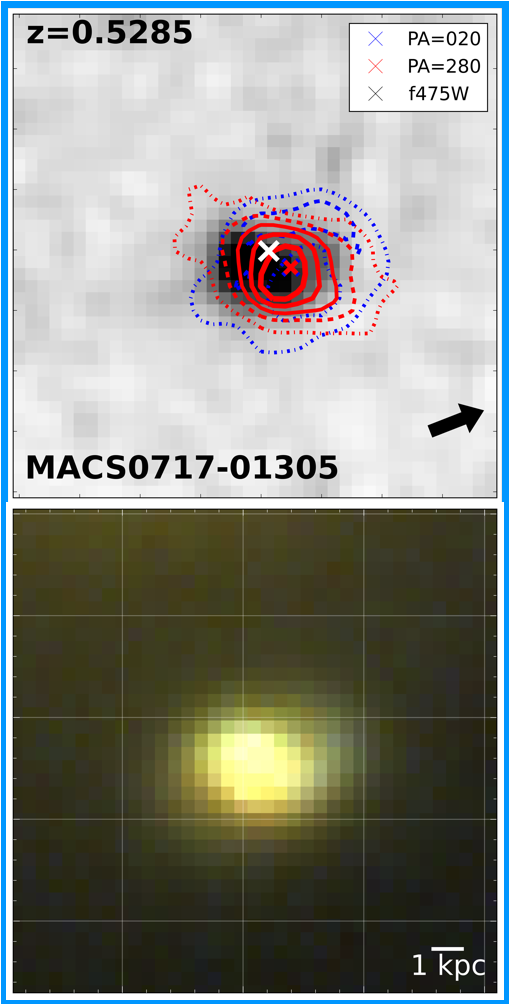}
  \includegraphics[scale=0.19]{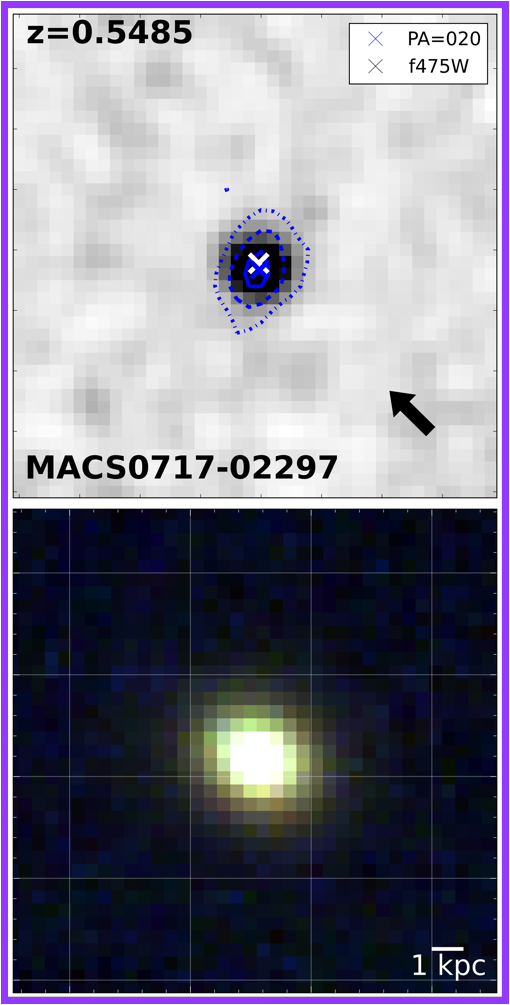}
  \includegraphics[scale=0.19]{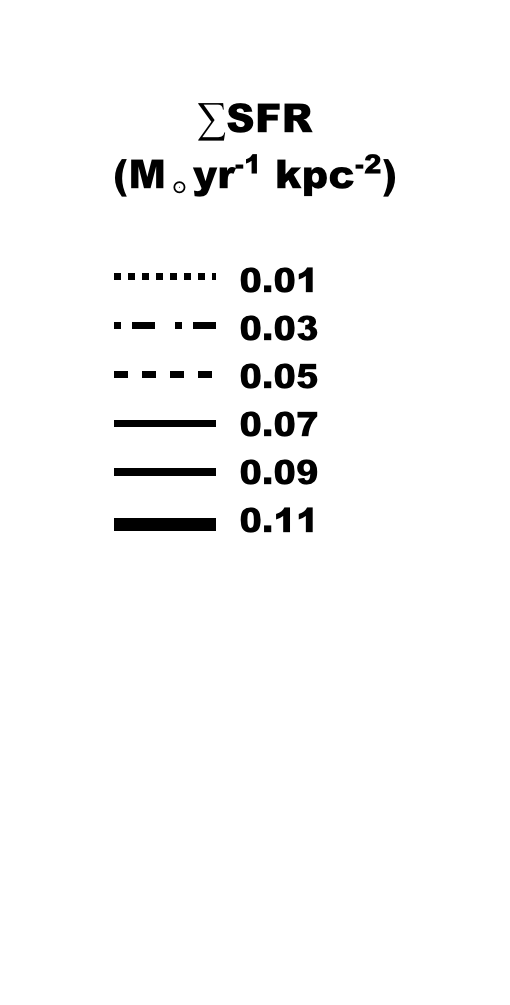}\\
\caption{Cluster galaxies with $0.8< \langle r($\Ha$)\rangle/ \langle r(F745W)\rangle <1.2$.  For each galaxy, in the upper panel the contour plots showing the \Ha maps superimposed on the image of the galaxy in the F475W filter are shown, in the bottom panel the color composite image of the galaxy based on the CLASH \citep{postman12} HST data is shown. The blue channel is composed by the F435W, F475W, F555W, F606W, and F625W (the last one only for \ma) filters, the green by the F775W, F814W, F850lp, F105W, F110W filters, and the red by the F125W, F140W, F160W filters. In the \Ha maps, different colors refer to the different PAs, Different line styles are indicated in the legend. A smoothing filter has been applied to the maps and an arbitrary stretch  to the images for display purposes.  for  Purple lines surround elongated galaxies (axis ratio in the  F475W filter $\neq1 \pm0.2$) light blue lines surround symmetric galaxies (axis ratio in the F475W filter $\sim1\pm0.2$). Arrows on the bottom right corner indicate the direction of the cluster center. The redshift of the galaxy is indicated on the top left corner.
\label{fig:c_size_Hec}}
\end{figure*}

\begin{figure*}
\centering
   \includegraphics[scale=0.19]{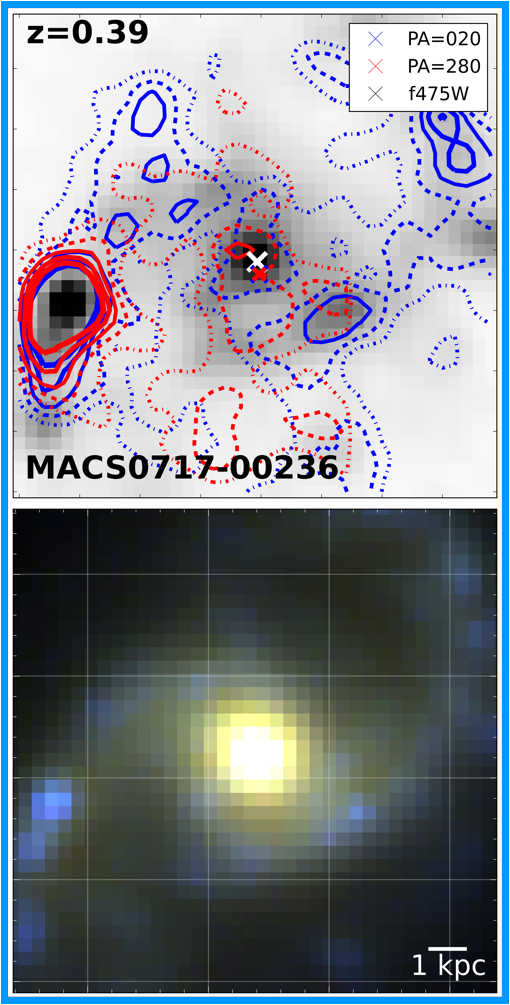}
   \includegraphics[scale=0.19]{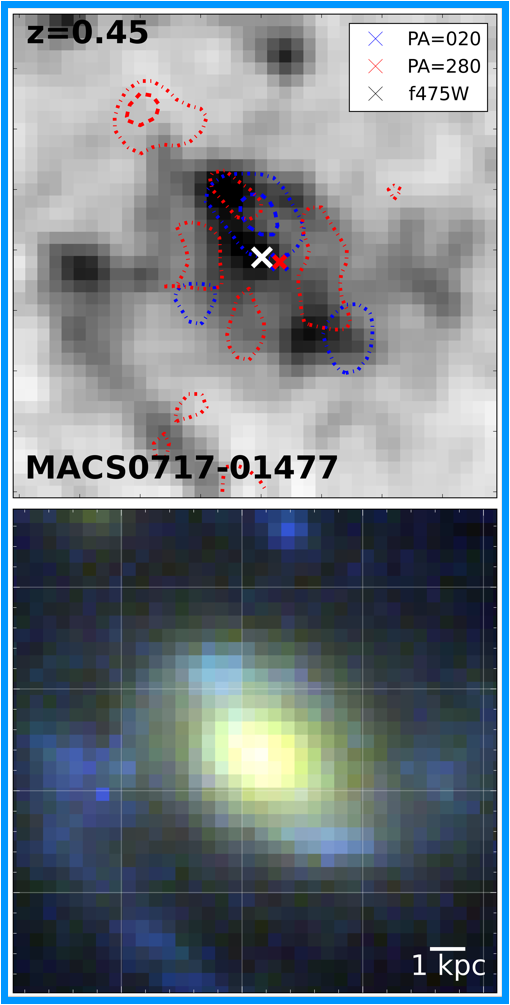}
   \includegraphics[scale=0.19]{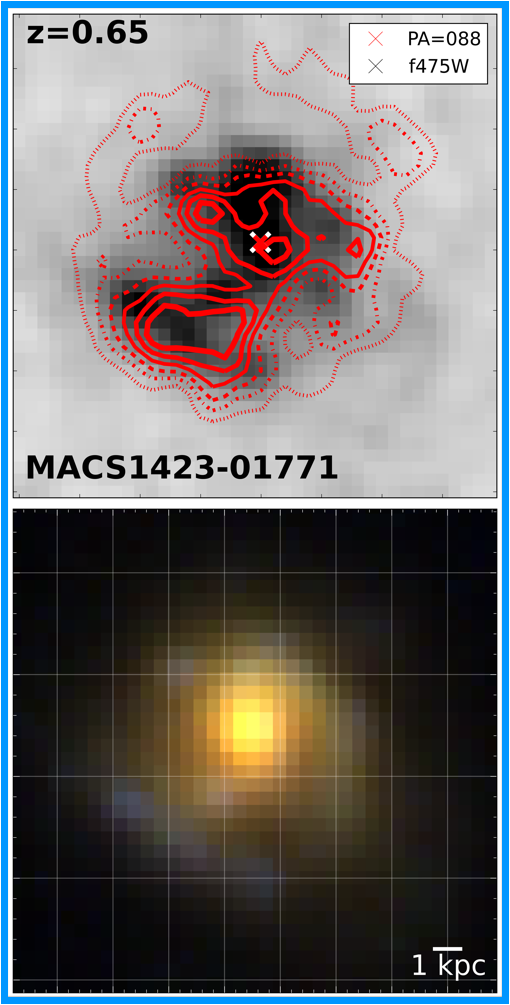}
\caption{Field galaxies with $0.8< \langle r($\Ha$)\rangle/ \langle r(F745W)\rangle <1.2$. Panels, lines and colors are as in Fig. \ref{fig:c_size_Hec}. 
\label{fig:f_size_Hec}}
\end{figure*}

\begin{figure*}
\centering
  \includegraphics[scale=0.19]{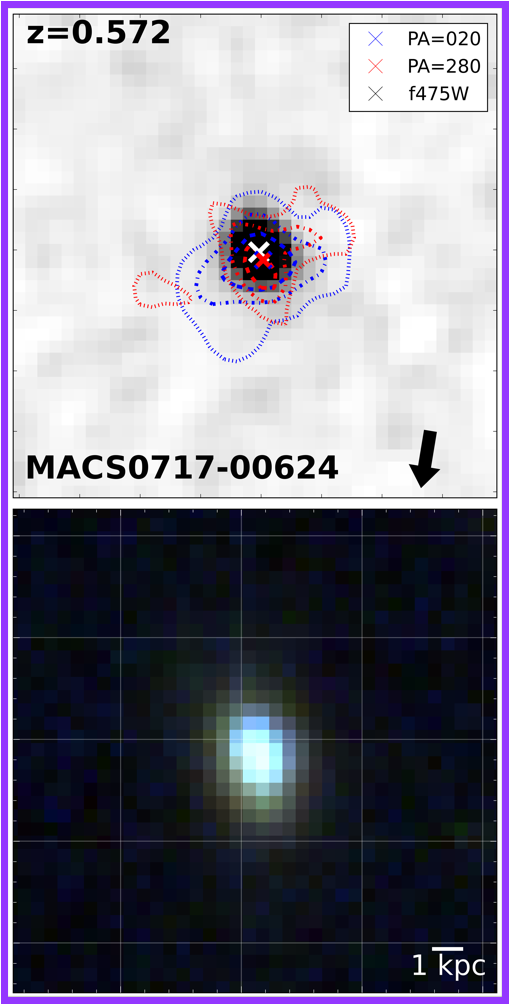}
  \includegraphics[scale=0.19]{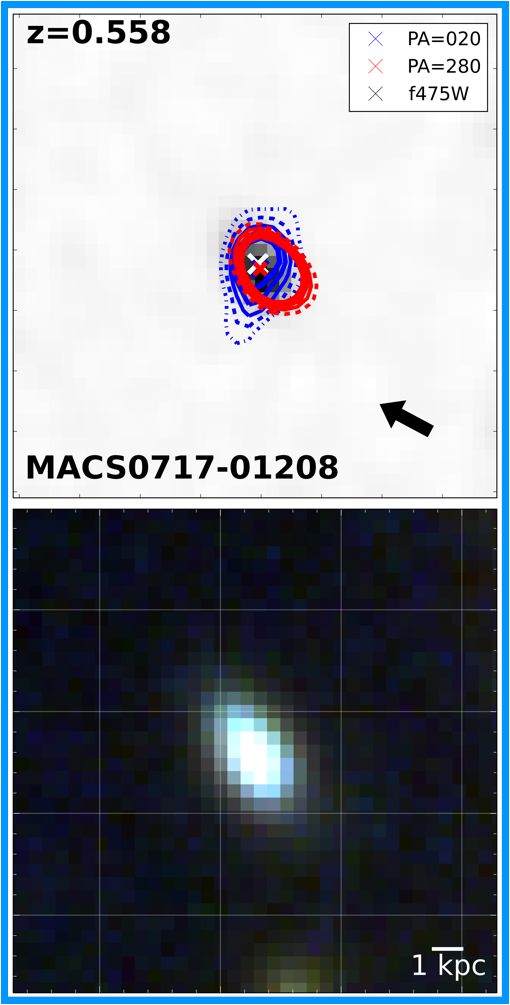}
  \includegraphics[scale=0.19]{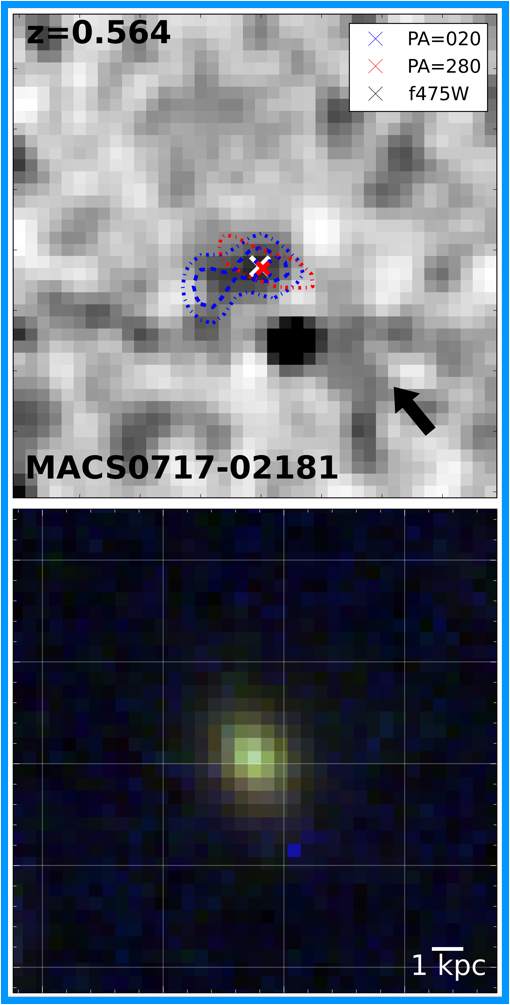}
  \includegraphics[scale=0.19]{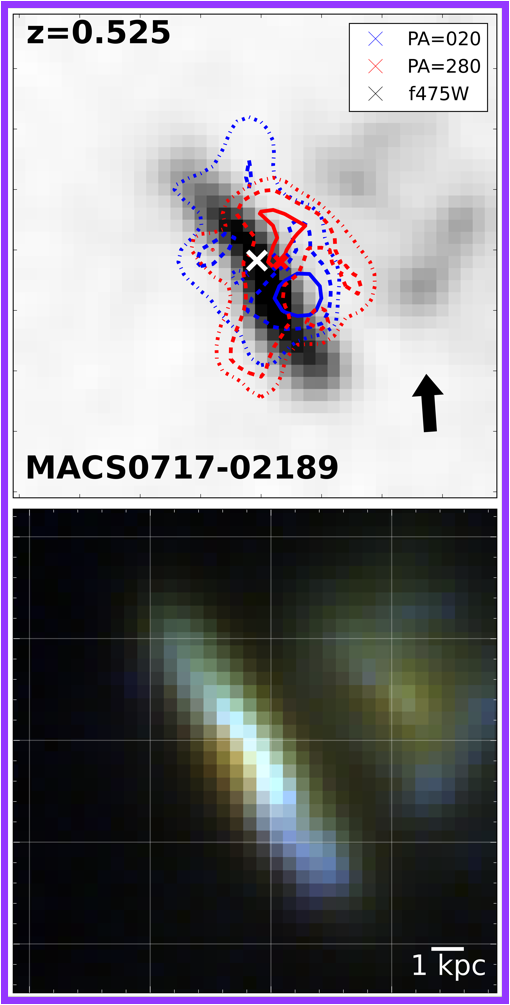}
  \includegraphics[scale=0.19]{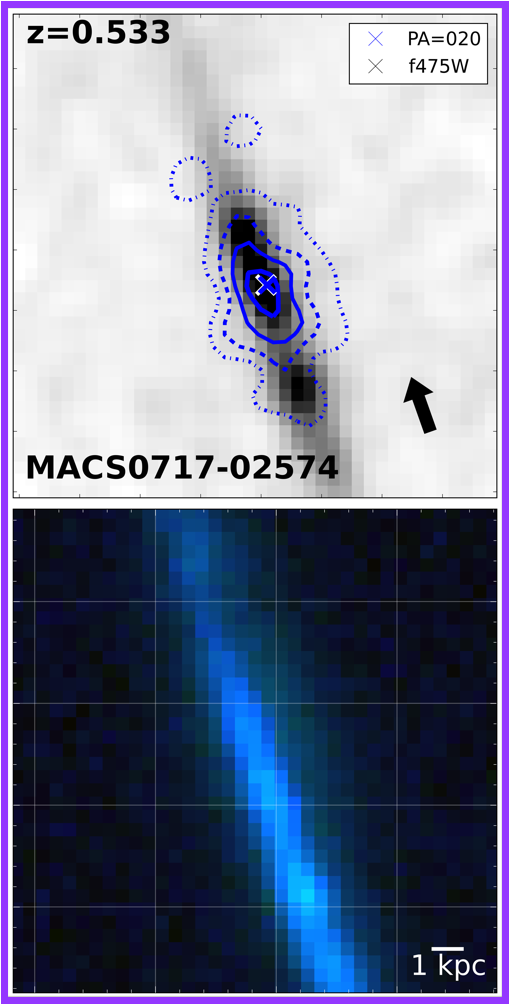}
  \includegraphics[scale=0.19]{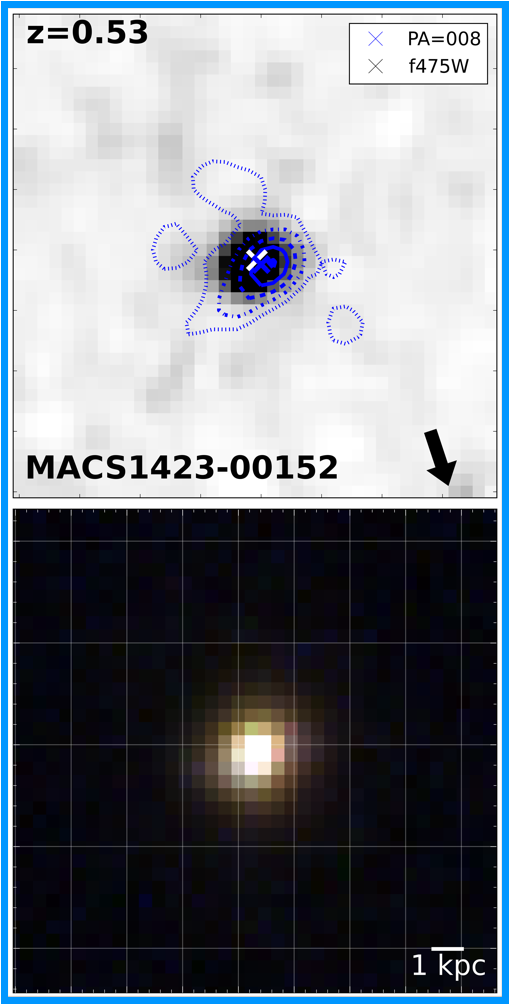}
  \includegraphics[scale=0.19]{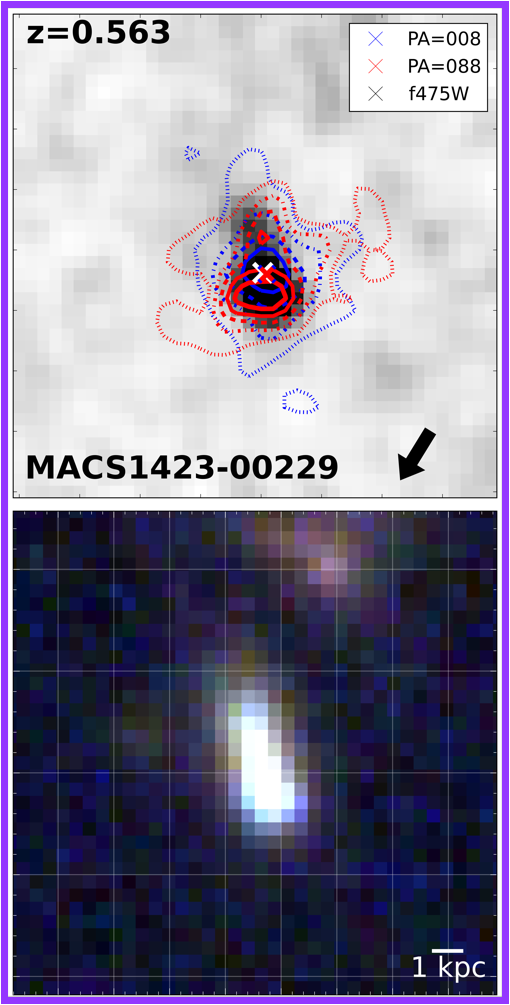}
  \includegraphics[scale=0.19]{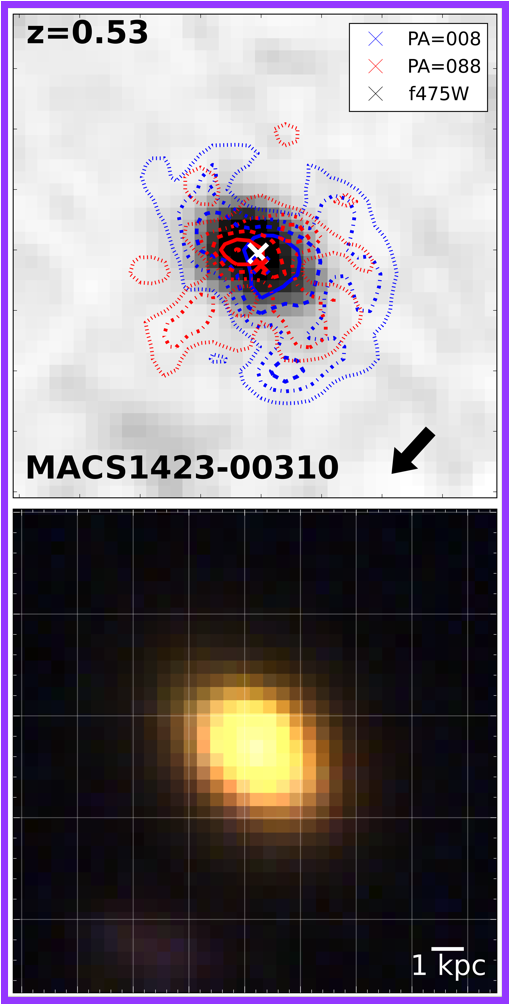}
  \includegraphics[scale=0.19]{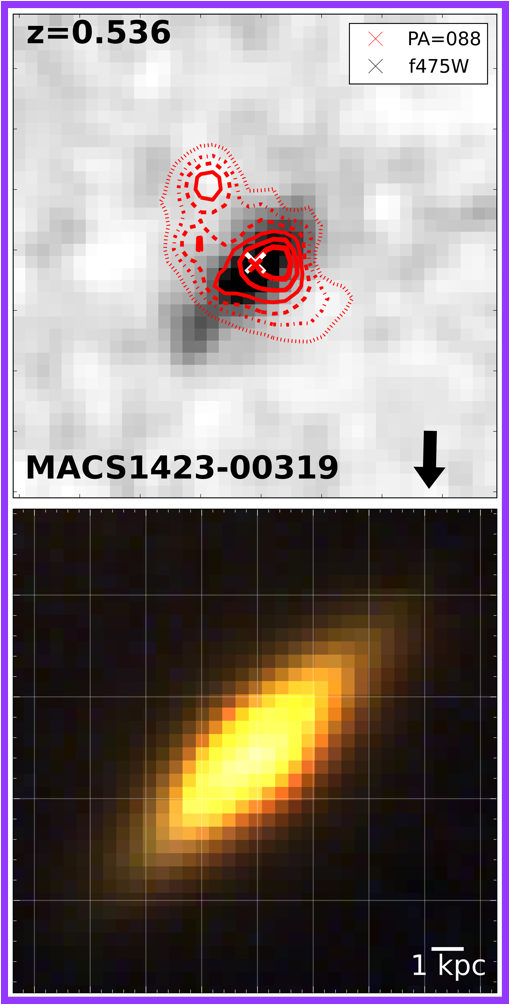}
  \includegraphics[scale=0.19]{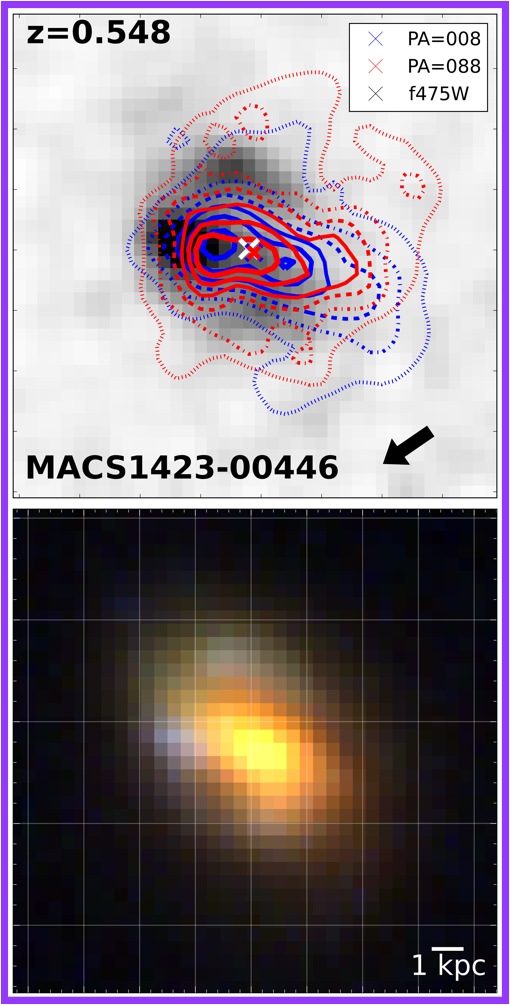}
  \includegraphics[scale=0.19]{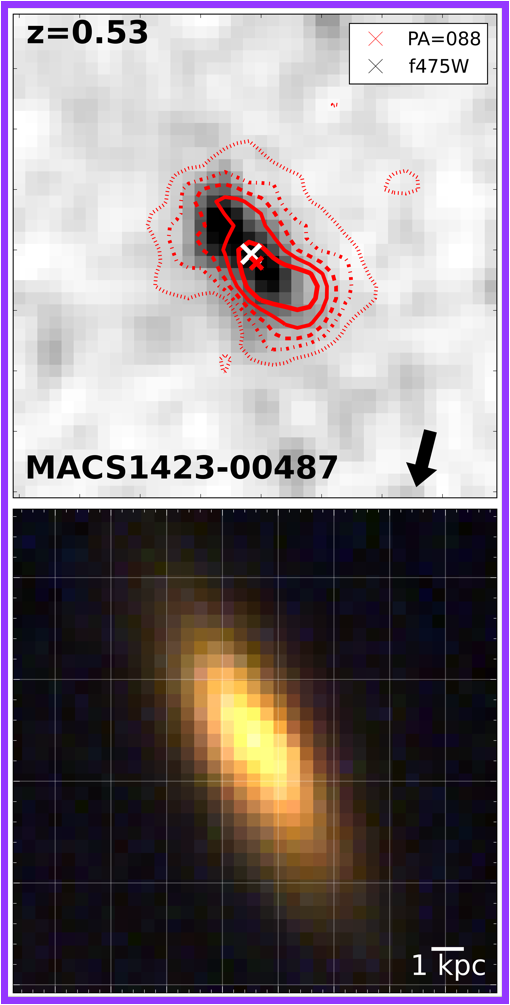}
  \includegraphics[scale=0.19]{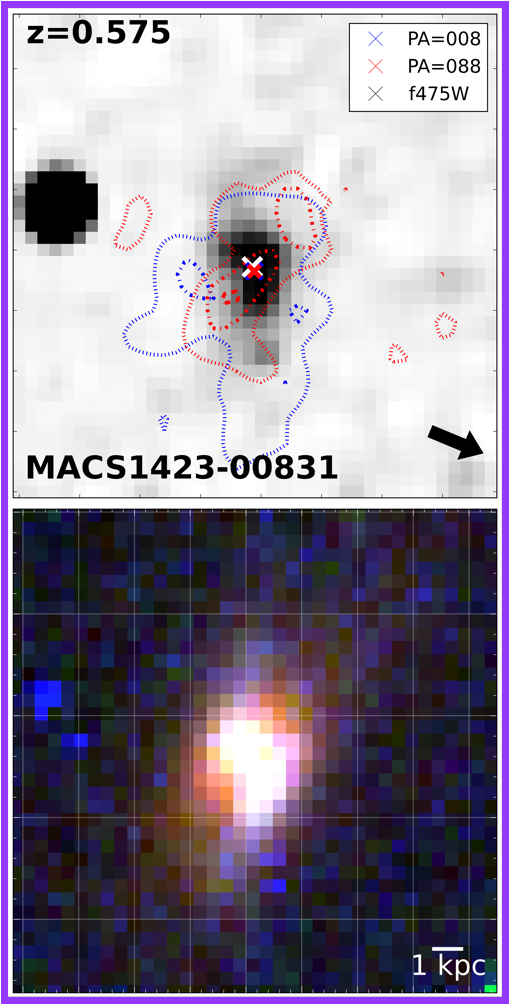}
  \includegraphics[scale=0.19]{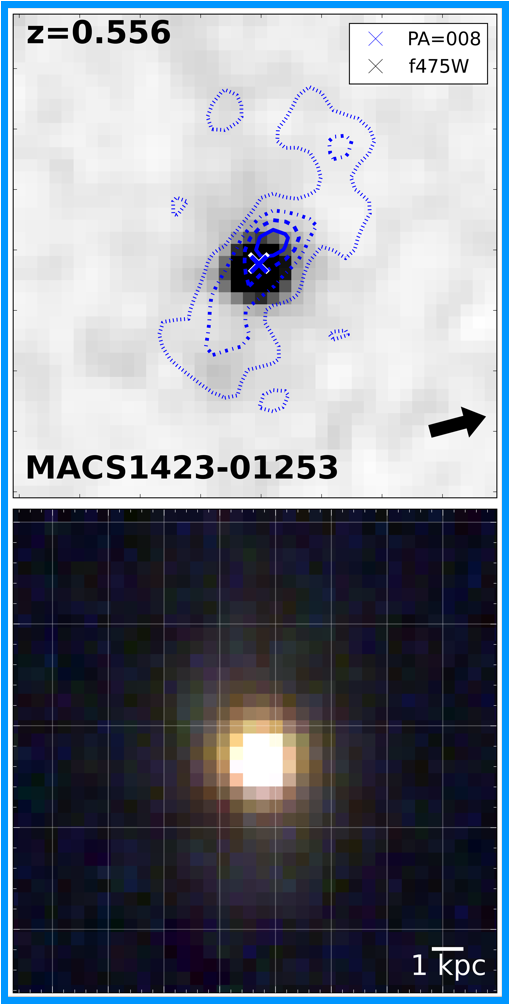}
  \includegraphics[scale=0.19]{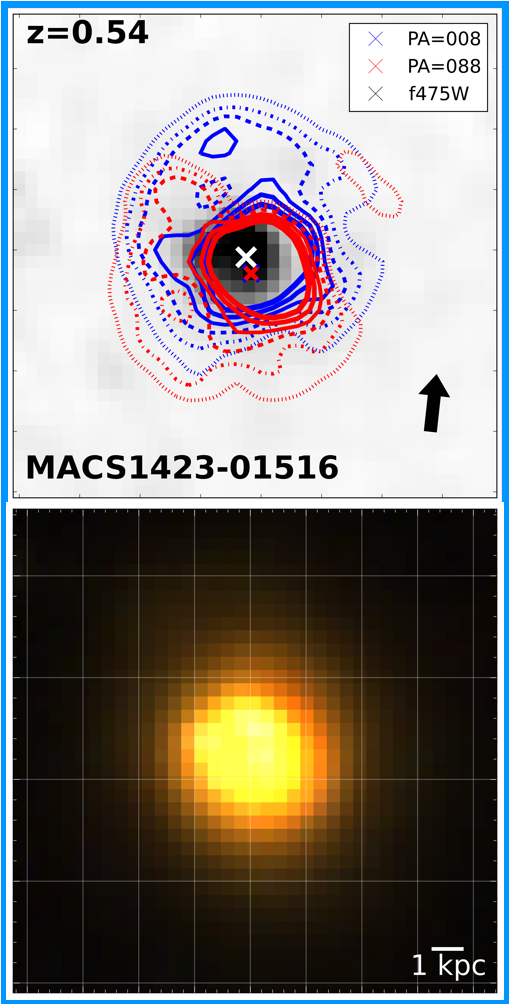}
  \includegraphics[scale=0.19]{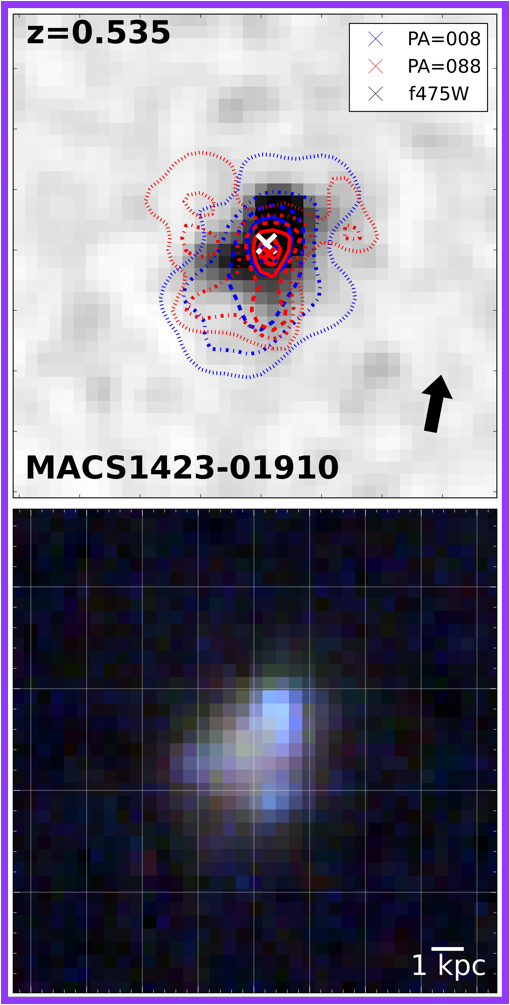}
\caption{Cluster galaxies with $ \langle r($\Ha$)\rangle/ \langle r(F745W)\rangle >1.2$. Panels, lines and colors are as in Fig. \ref{fig:c_size_Hec}. \label{fig:c_size_Hgc}}
\end{figure*}

\begin{figure*}[!t]
\centering
   \includegraphics[scale=0.19]{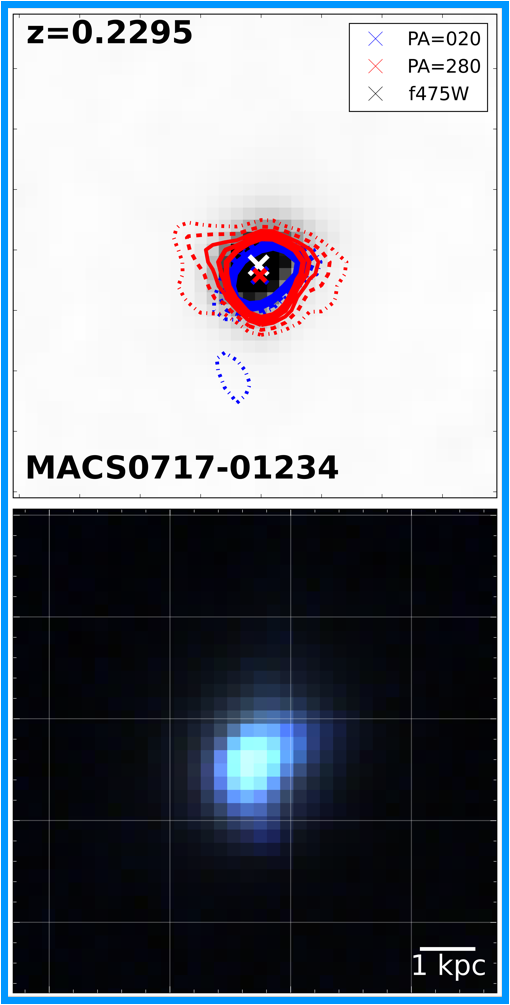}
   \includegraphics[scale=0.19]{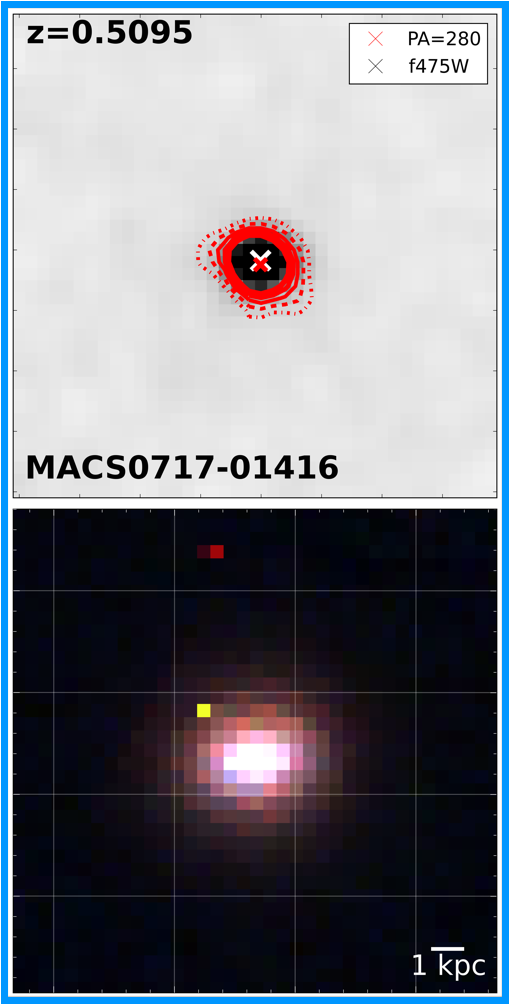}
   \includegraphics[scale=0.19]{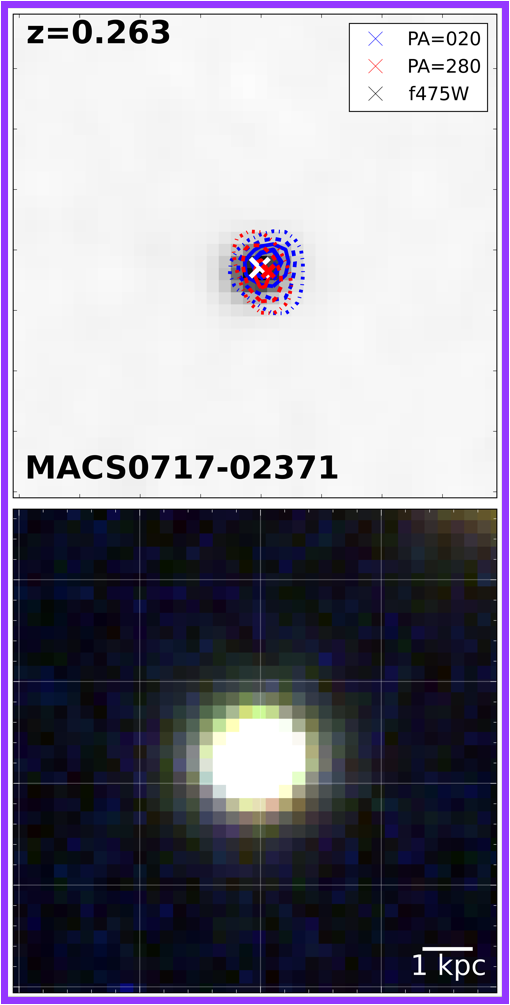}
   \includegraphics[scale=0.19]{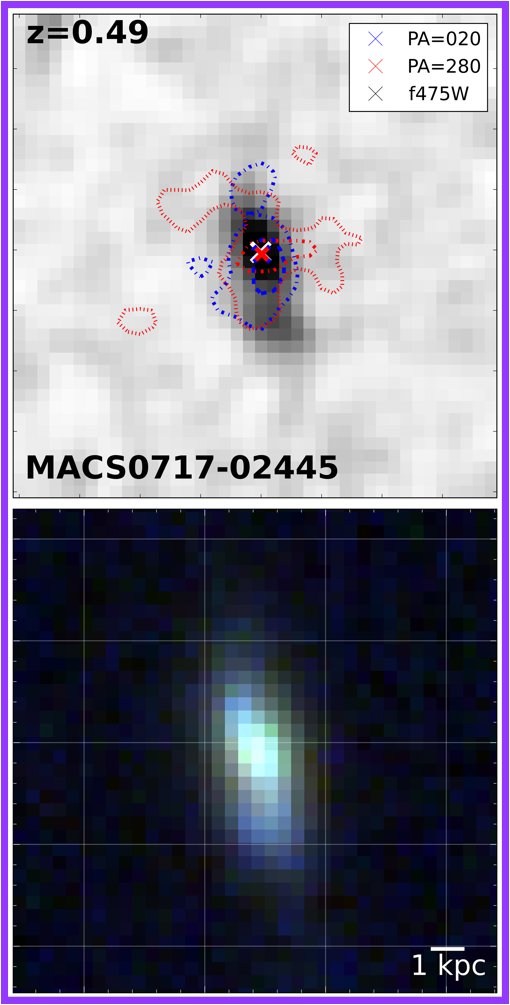}
   \includegraphics[scale=0.19]{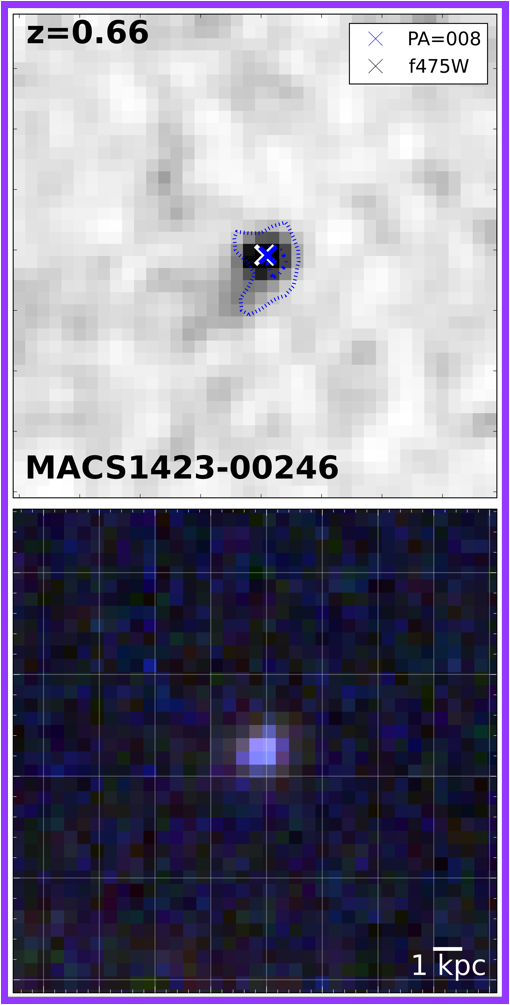}
   \includegraphics[scale=0.19]{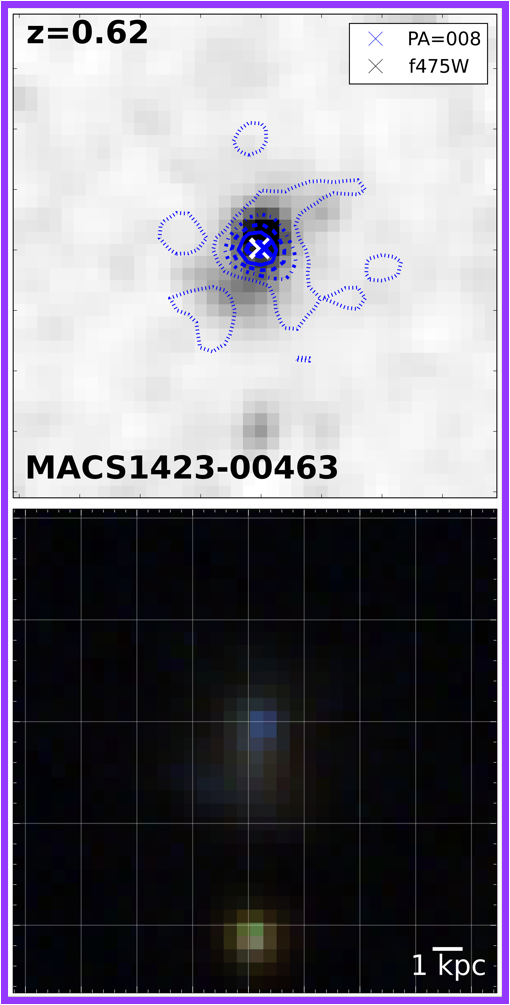}
   \includegraphics[scale=0.19]{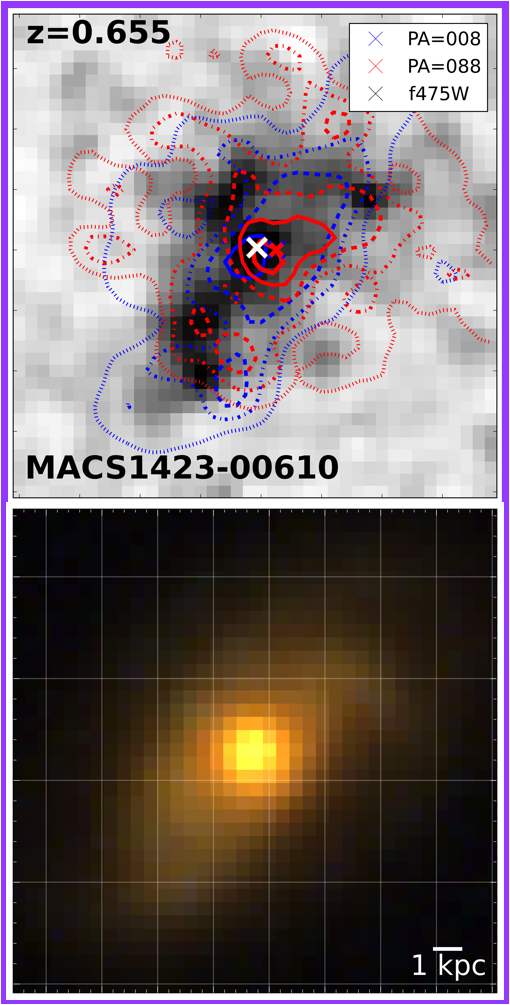}
   \includegraphics[scale=0.19]{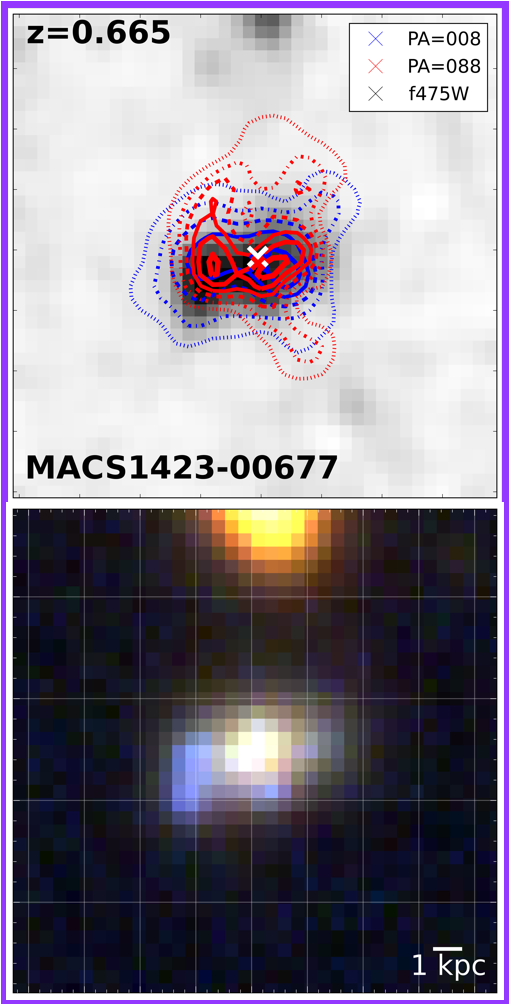}
   \includegraphics[scale=0.19]{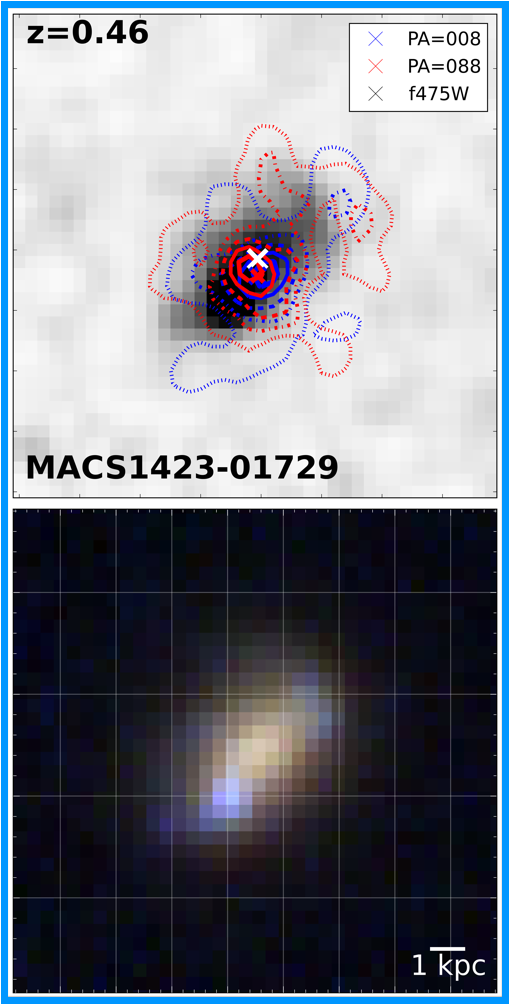}
   \includegraphics[scale=0.19]{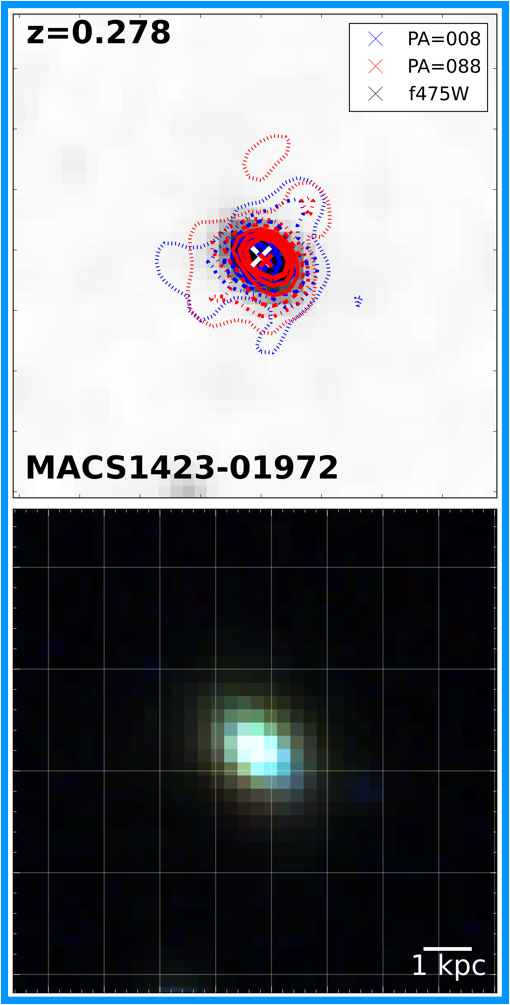}
 \caption{Field galaxies with $ \langle r($\Ha$)\rangle/ \langle r(F745W)\rangle >1.2$. Panels, lines and colors are as in Fig. \ref{fig:c_size_Hec}. 
\label{fig:f_size_Hgc}}
\end{figure*}

\begin{figure*}
\centering
   \includegraphics[scale=0.19]{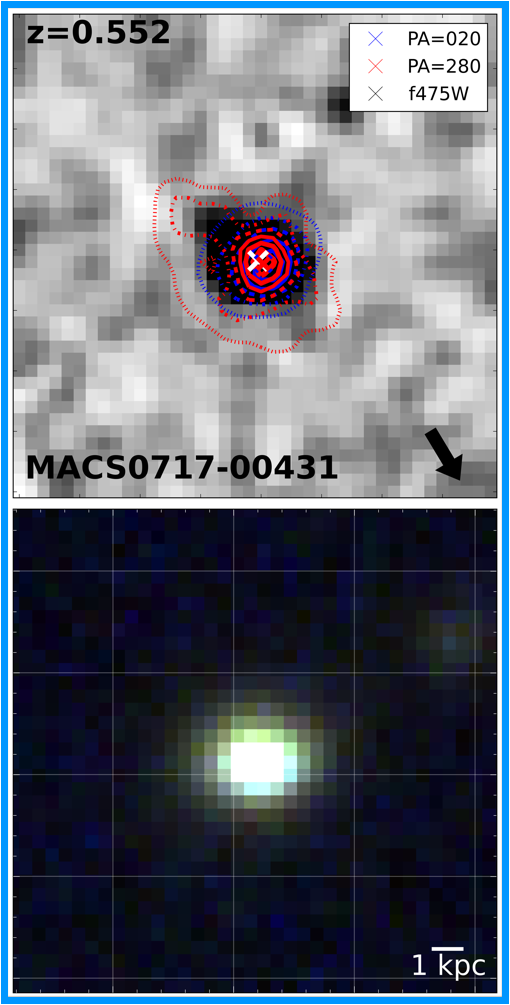}
   \includegraphics[scale=0.19]{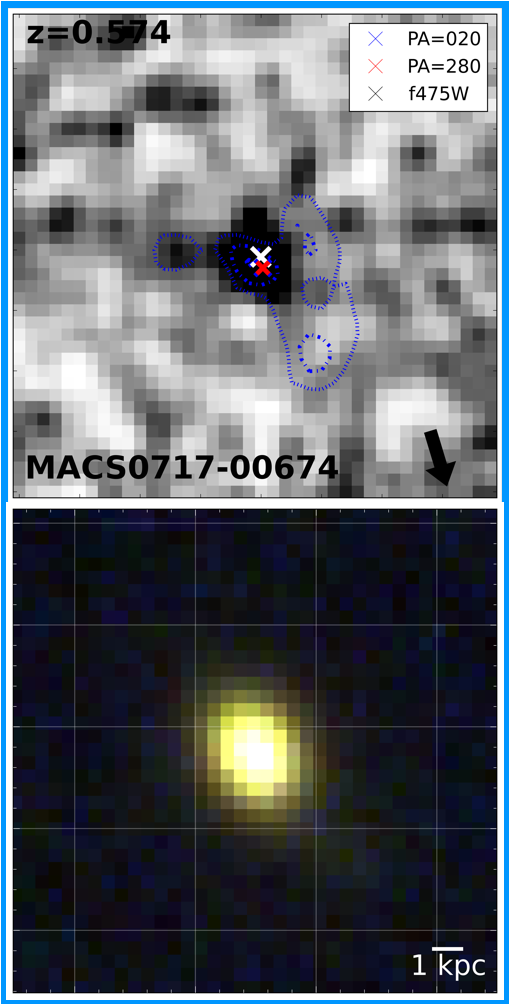}
   \includegraphics[scale=0.19]{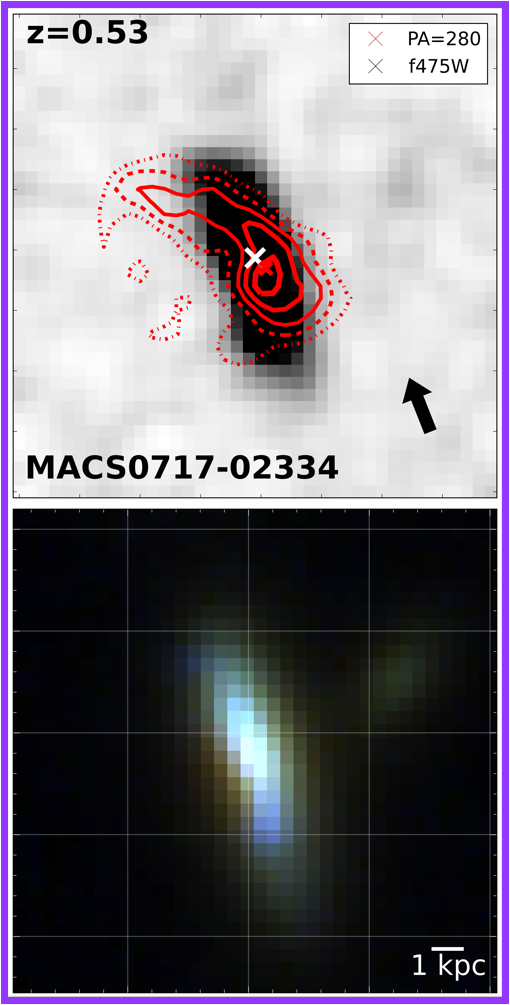}
  \includegraphics[scale=0.19]{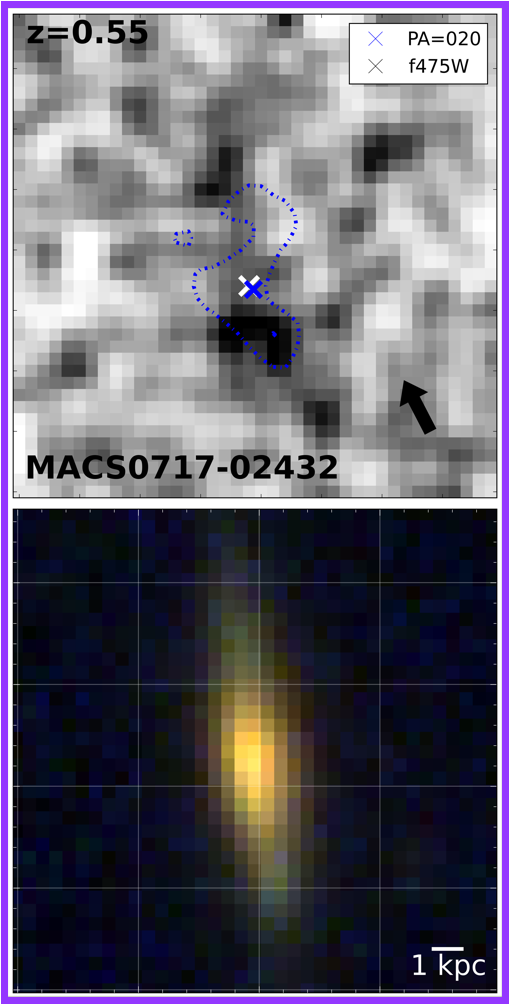}
\caption{Cluster galaxies with $ \langle r($\Ha$)\rangle/ \langle r(F745W)\rangle<0.8$. Panels, lines and colors are as in Fig. \ref{fig:c_size_Hec}. 
\label{fig:c_size_Hsc}}
\end{figure*}

\begin{figure*}
\centering
   \includegraphics[scale=0.19]{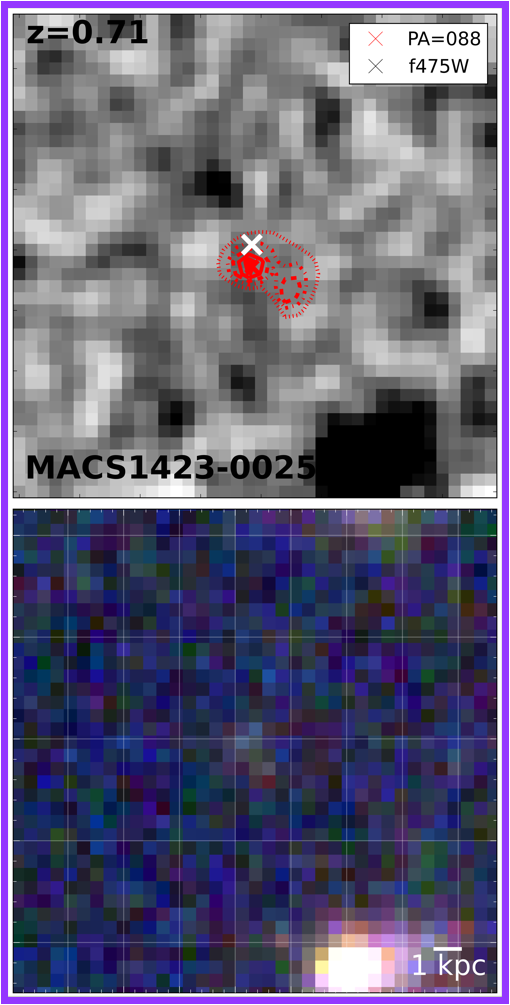}
   \includegraphics[scale=0.19]{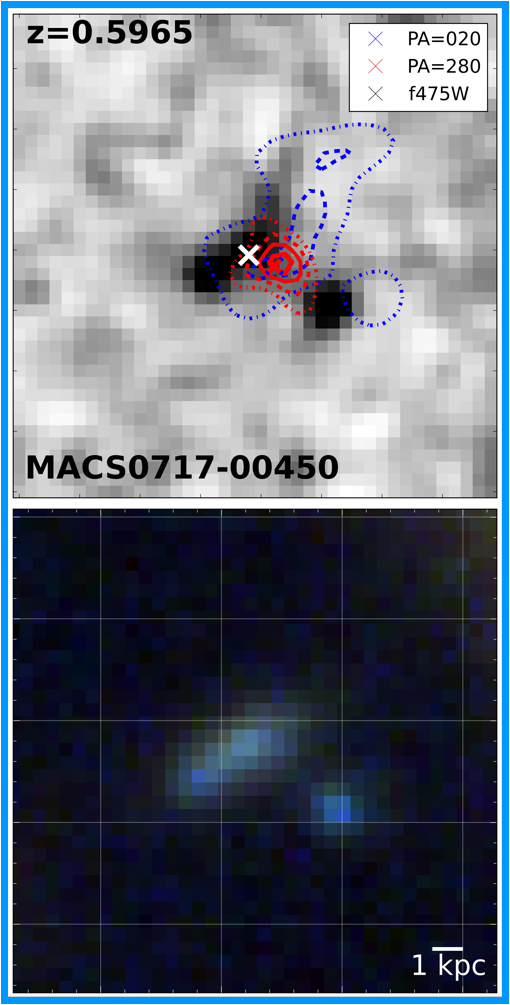}
   \includegraphics[scale=0.19]{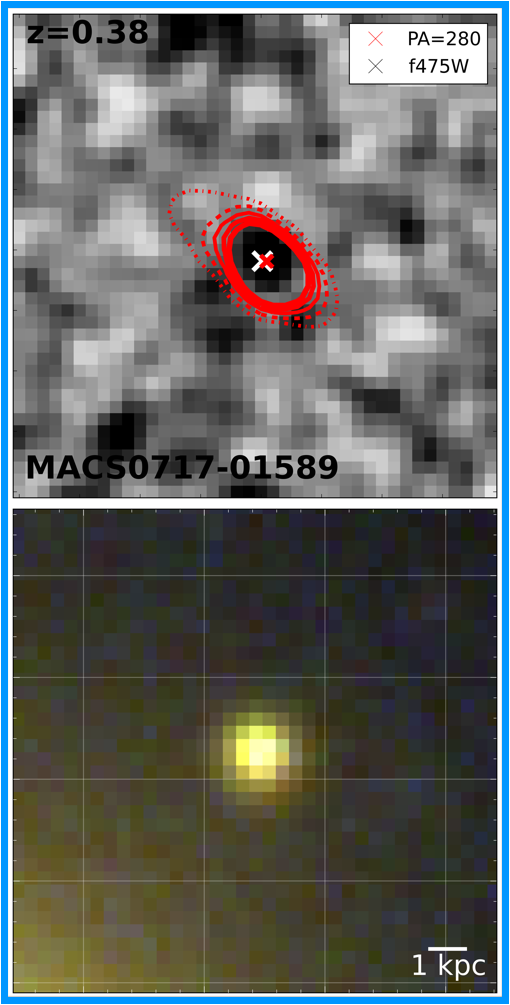}
   \includegraphics[scale=0.19]{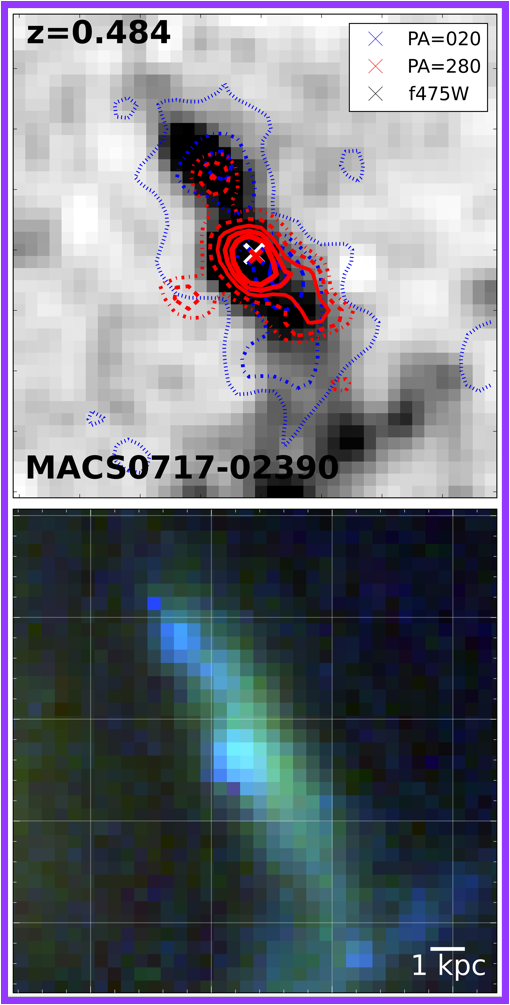}
 \caption{Field galaxies with $ \langle r($\Ha$)\rangle/ \langle r(F745W)\rangle <0.8$. Panels, lines and colors are as in Fig. \ref{fig:c_size_Hec}. 
\label{fig:f_size_Hsc}}
\end{figure*}

\begin{figure*}
\centering
\includegraphics[scale=0.65]{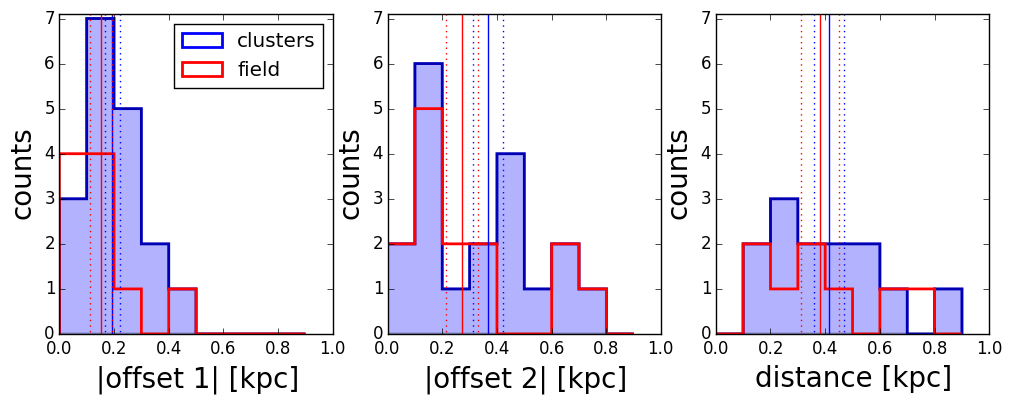}
\caption{Distribution of the offset of the \Ha emission with respect to the continuum. Left and central panel are the offsets in the two direction separately, right panel shows the combined offset, for the galaxies with both PAs.  Vertical solid and dotted lines represent the means with errors. Most of the galaxies present an offset between the two emission. The displacement is slightly larger in clusters than in the field.
\label{fig:offset}}
\end{figure*}

We  group galaxies according to 1) the ratio of the  average \Ha size  to the average size of the rest-frame UV continuum shown in the top right panel of Fig. \ref{fig:size}, and  2) the axis ratio of the continuum. 
The first classification scheme states whether  the current star formation is occurring at larger or smaller radii than the recent star formation. We note that similar results are obtained when we consider the rest-frame optical continuum, which traces the older  stars in the galaxy. The second  
is a rough attempt to describe the galaxy morphology in the light of the continuum. 
However, we note that all these classes of objects contain very heterogeneous cases with a variety of different features. It is therefore quite hard to perform a strict  classification.

Figures 5-10 
show the \Ha maps obtained as described in  Section 3, for all galaxies in our sample. For each galaxy, also a  color composite image of the galaxy based on the CLASH \citep{postman12} HST data is shown. The blue channel is composed by the F435W, F475W, F555W, F606W, and F625W (the last one only for \ma) filters, the green by the F775W, F814W, F850lp, F105W, F110W filters, and the red by the F125W, F140W, F160W filters.

Figures \ref{fig:c_size_Hec} and \ref{fig:f_size_Hec} show the 6/25 cluster and 3/17 field galaxies with similar sizes in  \Ha  and in the rest-frame UV continuum ($0.8< \langle r($\Ha$)\rangle/ \langle r(F745W)\rangle <1.2$). 4 cluster galaxies show elongated sizes in the light of the rest-frame UV continuum (axis ratio >1.2), while all field galaxies show symmetric shapes. Nonetheless, 3 galaxies in the field have clearly spiral morphologies. 

Figures \ref{fig:c_size_Hgc} and \ref{fig:f_size_Hgc} show all galaxies with size of the \Ha light larger than the size measured from the continuum ($ \langle r($\Ha$)\rangle/ \langle r(F745W)\rangle >1.2$),  in clusters and in the field respectively. The great majority of cluster and field galaxies fall into this class (15/25 and 10/17 respectively). Of these, 11 in clusters  and 6 in the field show an elongated shape.
Though being star forming, most of these galaxies show an early-type morphology in the color composite images. In clusters, this might be a sign of ongoing stripping.

Few galaxies have \Ha sizes smaller than continuum sizes ($ \langle r($\Ha$)\rangle/ \langle r(F745W)\rangle <0.8$)  and are shown in Figures \ref{fig:c_size_Hsc} and \ref{fig:f_size_Hsc}. In both environments, 2 out of 4 galaxies  show  symmetric profiles. 

In general, our sample includes galaxies with a variety of morphologies and we find that there is no clear correlation between the extent of the \Ha emission and the galaxy color or morphology in the color images. This might suggest that there is no a unique mechanism responsible for extension of the \Ha, but that different processes might be at work  in galaxies of different types. 

Overall, both in clusters and in the field 60\% of  galaxies show \Ha emission more extended than the emission in the rest-frame UV  continuum. Half of the galaxies in the field show a symmetric shape, 35\% in clusters. 

When comparing the maps at different wavelengths, we also observe that the peak of the \Ha emission is displaced with respect to the  F475W continuum emission. Figure \ref{fig:offset} shows the distribution of the absolute value of the offsets in the two directions (obtained from the two different PAs) and, for the galaxies with both PAs, the real distance between the two peaks, obtained by combining the offsets. In both environments, the displacement is always smaller than 1 kpc. There is no a preferential direction of the offset. There are hints that galaxies in clusters are characterized by a marginally larger offset than field galaxies, but a larger number statistics will be needed to confirm the trends. The existence of the offset suggests that in most galaxies the bulk of the star formation is not occurring in the galaxy cores.
Unsurprisingly given the small sample statistics, cluster and field means are compatible within the errors and a K-S test can not reject the hypothesis that the two distributions are drawn from the same parent distribution.   

 We note that in our analysis we have made the assumption that there is no spatial variation in extinction across the galaxy. Nonetheless, high-resolution imaging in multiple HST bands \citep{wuyts12} as well as analysis of such data in combination with \Ha maps extracted from grism spectroscopy \citep{wuyts13} indicate that such an assumption may be over-simplistic, particularly in the more massive galaxies where the largest spatial color variations are seen. It is hard to anticipate how corrections for non-uniform extinction might affect our conclusions, since the correction to the sizes will depend on the actual distribution of dust. For example, if dust is mostly in the centers (like a dust lane), it would make us overestimate the F475W sizes more than the \Ha sizes. Conversely, if dust is mostly in the outskirts the correction could lead us in the opposite direction. Reaching a firm conclusion would require obtaining Hb maps to trace the Balmer decrement. Unfortunately H$_\beta$ is too blue for the G102 grism for these clusters. 

\subsection{Maps of \Ha and position within the clusters}

\begin{figure*}
\centering
\includegraphics[scale=0.35]{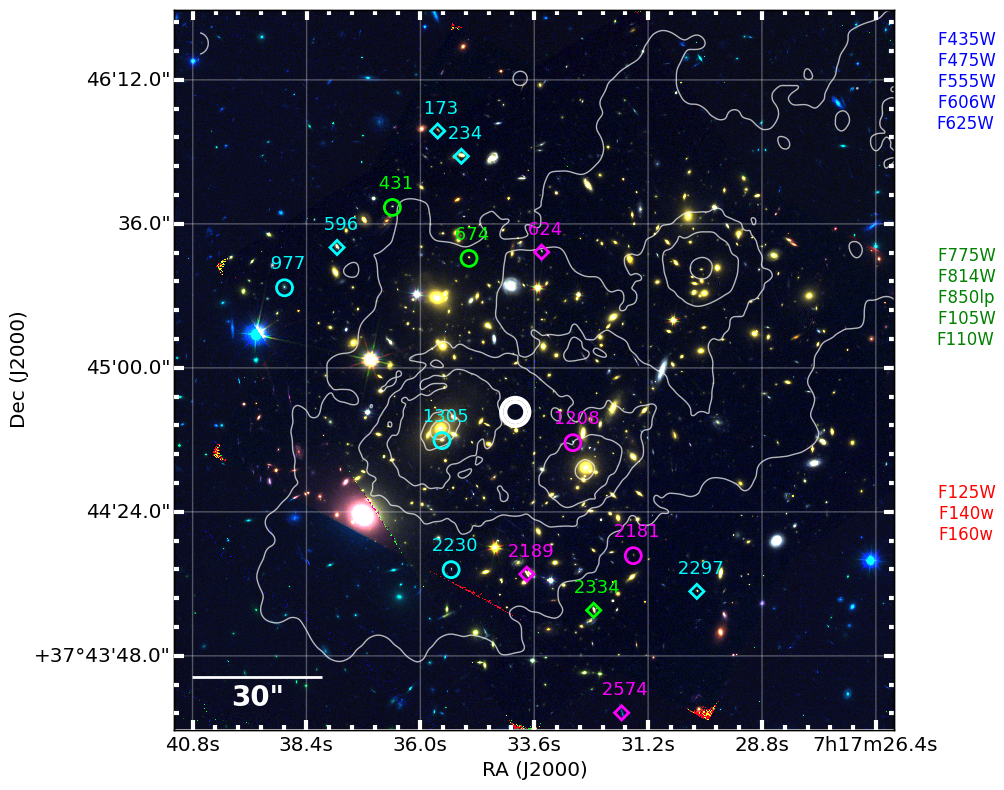}
\includegraphics[scale=0.35]{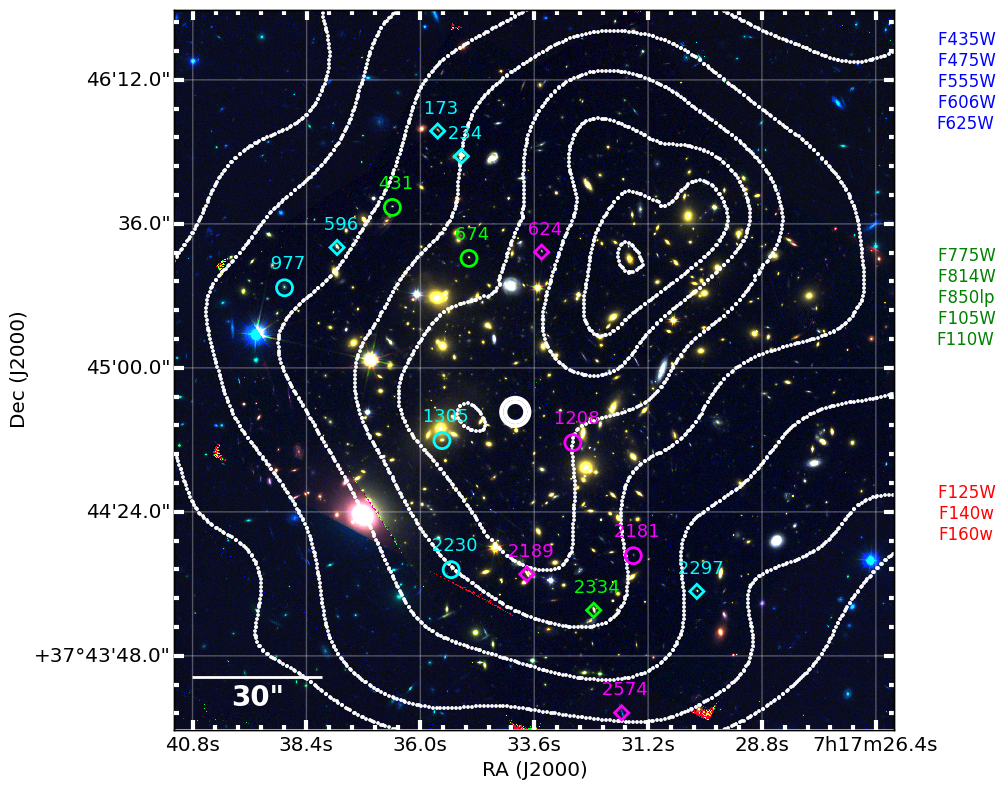}
\includegraphics[scale=0.35]{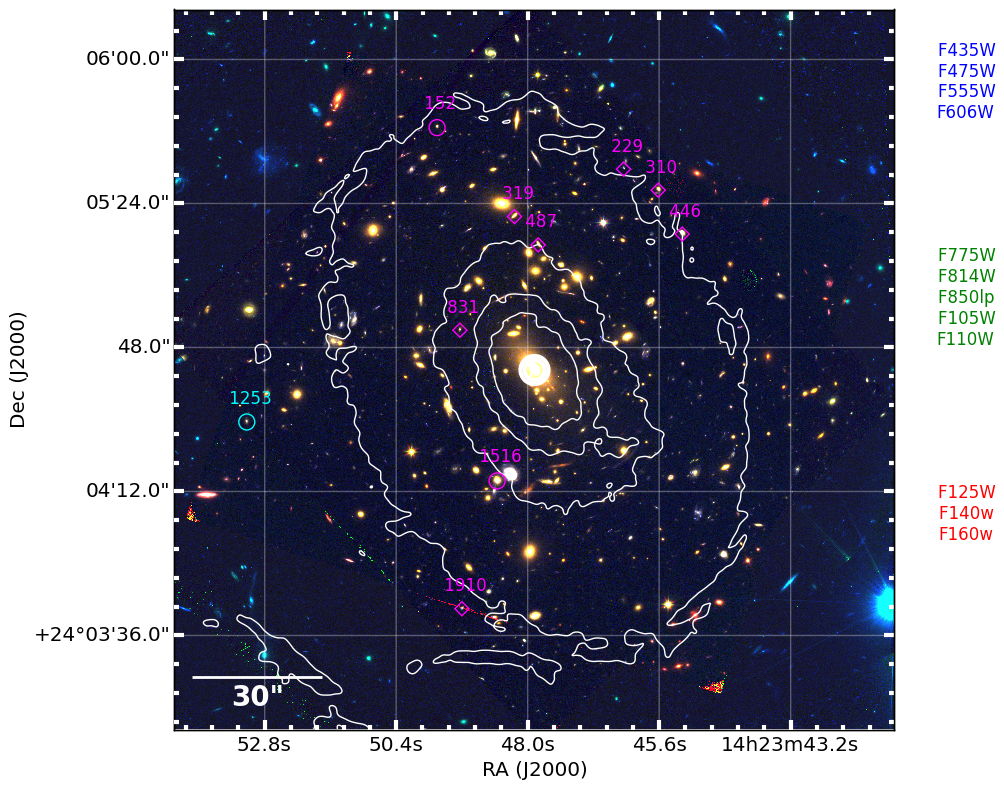}
\includegraphics[scale=0.35]{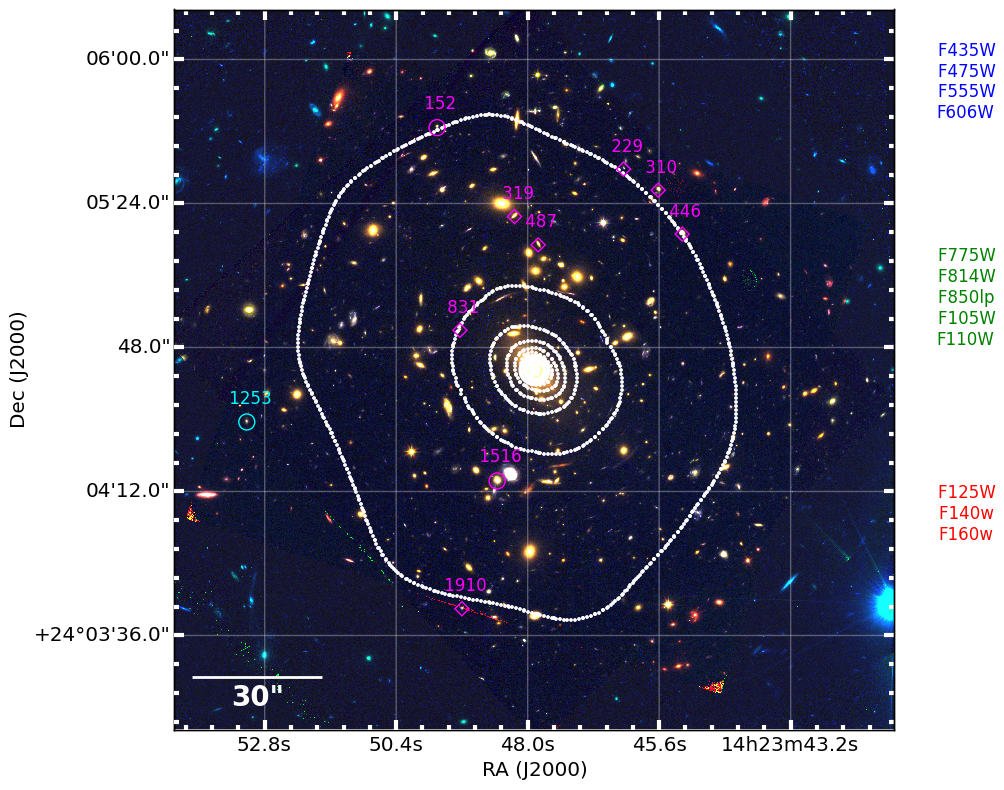}
\caption{Color composite image of \ma (upper panels) and \mb  (bottom panels) based on the CLASH \citep{postman12} HST data. The blue, green, and red channels are composed by the filters on the right.  Galaxies with different size in \Ha and continuum are shown with different colors. Cyan: size(\Ha)=size(continuum),  magenta: size(\Ha)>size(continuum), green: size(\Ha)<size(continuum).  Galaxies with different size ratios in the continuum  are shown with different shapes. Diamonds: elongated continuum,  circles: symmetric continuum.
The white thick circle represents the cluster center. In the left panel, surface mass density contours are  overplotted. A smoothing filter has been applied to the images for display purposes.   Contours show   $(0.5,1, 1.5, 2, 5, 10) \times 10^{10}\textrm{M}_\odot \,  \textrm{kpc}^{-2}$. In right panel, X-ray emission is  overplotted. Contours are  spaced on a "sqrt" scale from  0.0002 to 0.004 counts/s/px  for \ma and from 0.0005 to 0.2 counts/s/px for \mb. 
\label{fig:image}}
\end{figure*}

For cluster galaxies, we can correlate their morphology with their location in the cluster,  the surface mass density distribution and  the X-ray emission, as shown in Fig. \ref{fig:image}.
 
The mass maps were produced using SWUnited reconstruction code described in detail in \citet{bradac04b}
and \citet{bradac09}.  The method uses both  strong and weak lensing mass reconstruction on a non-uniform adapted grid. From the set of potential values we determine all observables (and mass distribution) using derivatives.  The potential is reconstructed by maximizing a log likelihood which uses image positions of multiply imaged sources, weak lensing ellipticities, and regularization as constraints. For both MACS1423 and MACS0717 we use CLASH data. In addition, for MACS0717 we make use of the arcs identified in HST archival imaging prior to Hubble Frontier Fields \citep{zitrin09,limousin10} and spectroscopic redshifts obtained by \citet{ma09, limousin12,ebeling14}. For MACS1423 we use spectroscopic redshifts and images identified in \citet{limousin10}; and we add additional multiply imaged systems discovered by our team. 

The X-ray images are based on Chandra data, and are described in \cite{mantz10} and \cite{vonderlinden14a}.  For the contours shown in Fig.\ref{fig:image}, the images have been adaptively smoothed, after removing point sources identified in \cite{ehlert13}.

In both clusters, galaxies are located within $\sim$0.4r$_{500}$ and do not seem to avoid the cluster cores, even though there might be  possible projection effects. The two clusters present very different surface  mass density distribution and X-ray emission: while \ma extends along the north-west - south-east direction and has more than one main peak in the emissions, \mb shows a symmetric surface mass density distribution and X-ray emission. We note that \mb passes the very strict requirements on relaxedness defined in \cite{mantz14}. In  \ma we find galaxies with both truncated or extended \Ha with respect to the rest-frame UV continuum, in \mb all galaxies 
have  \Ha light more extended than the continuum light. 
Despite the small number statistics, it appears  that the truncated objects are only found between the merging clusters, suggesting that the spatial distribution of \Ha is indeed related to cluster dynamics:  the most extreme cases of stripping are expected to take place in interacting systems \citep[e.g.,][]{owen06, smith10, owers12}.

\begin{figure*}
\centering
\includegraphics[scale=0.45]{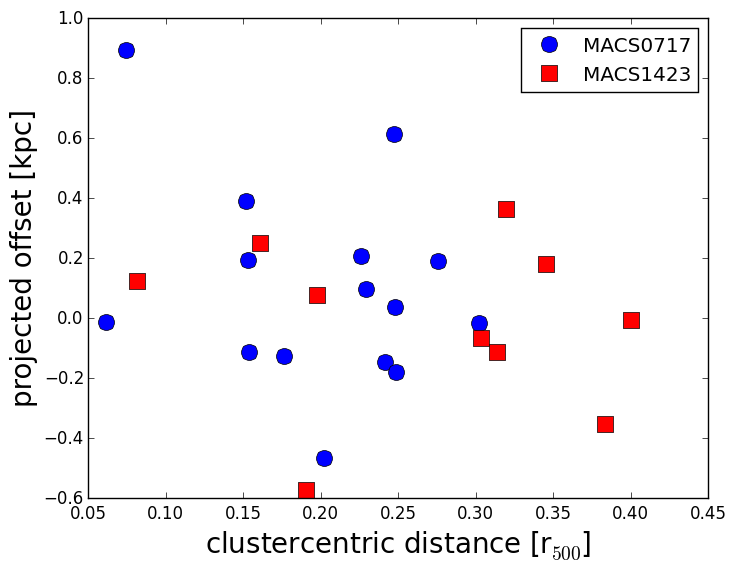}
\includegraphics[scale=0.45]{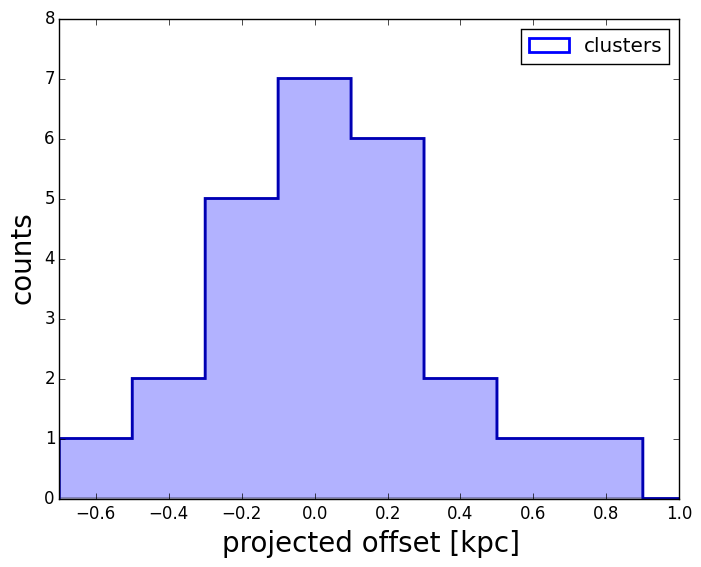}
\caption{Left: Projected offset along the cluster radial direction - distance from the cluster center relation for galaxies in clusters. Right: Projected offset along the line of sight distribution.
No trends with distance are detected, indicating that the cluster center is not affecting the position of the peak of the \Ha emission.
\label{fig:dist_offset}}
\end{figure*}

Figure \ref{fig:dist_offset}  quantifies the relation between  the projected offset along the  cluster radial direction and the distance of the galaxy from the cluster center, for cluster members. While most of the galaxies have a projected offset within $\pm$0.2 kpc, there are some galaxies with a larger offset. Almost half of the galaxies (55\%) have a positive offset, the other half have a negative offset. No trends with distance are detected, indicating that the cluster center is not affecting the  position of the peak of the \Ha emission.
Galaxies with different \Ha extension are not clustered in particular regions of the clusters.

\begin{figure}
\centering
\includegraphics[scale=0.23]{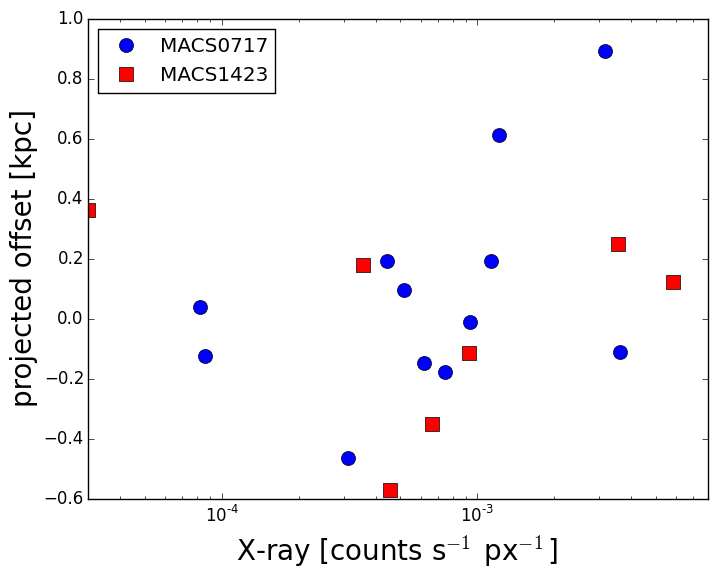}
\includegraphics[scale=0.23]{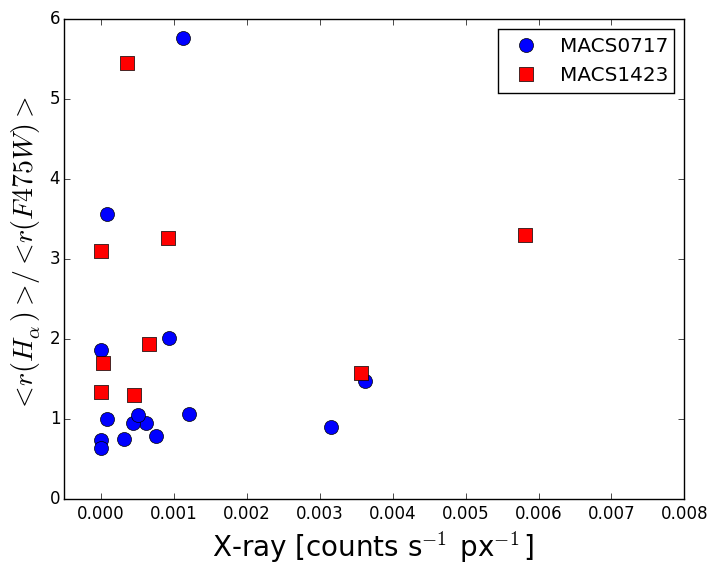}
\includegraphics[scale=0.23]{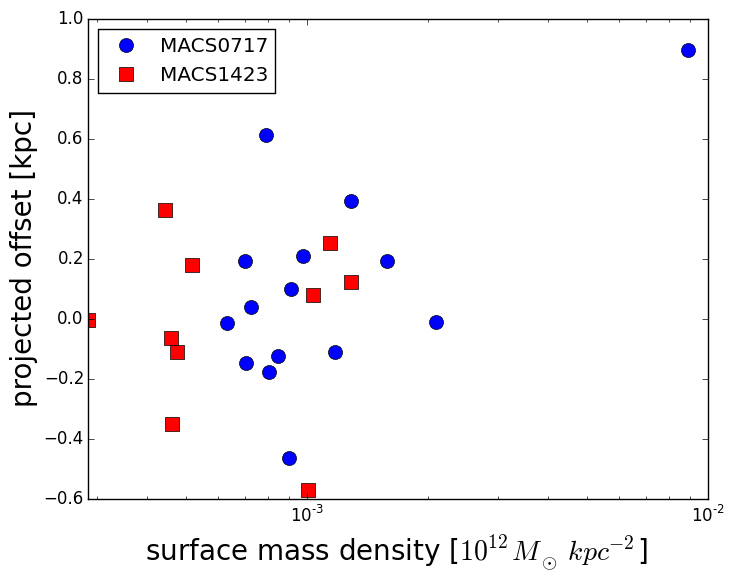}
\includegraphics[scale=0.23]{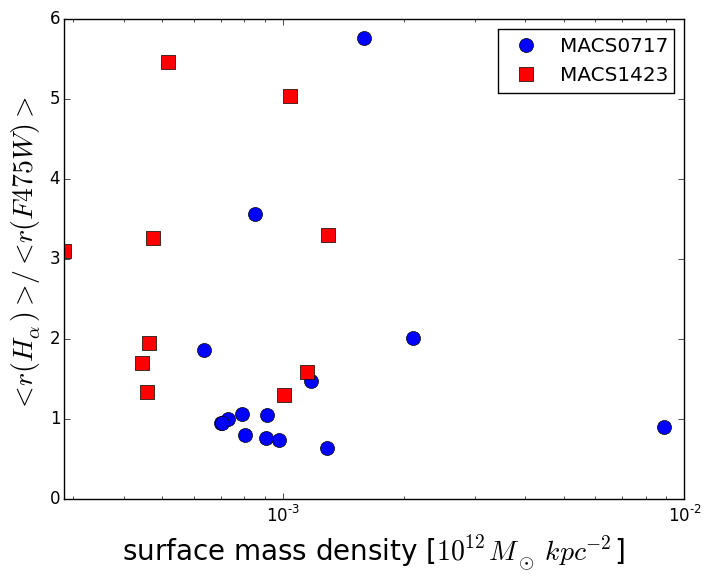}
\caption{Galaxy properties - cluster properties correlations. Upper row: X-ray emission, bottom row: Surface mass density. Left panels: projected offset along the line connecting the peak of the emission in the continuum and the cluster center. Right panels: ratio of r(\Ha) to r(F745W).   Shocks and strong gradients in the X-ray IGM might alter the relative position between the peak of the \Ha emission and the peak of the light of the of the young stellar population ($\sim$100 Myr), even though there is not statistically significant evidence to support this conclusion. Shocks and  gradients do not alter the relative extension of the \Ha with respect to the continuum light. 
\label{fig:corr}}
\end{figure}
There is increasing evidence for a correlation between the efficiency of the stripping phenomenon and the presence of shocks and strong gradients in the X-ray IGM \citep[e.g.,][]{owers12, vijayaraghavan13}.  Indeed,  Figure \ref{fig:corr} hints at potential correlations between the offset and  X-ray emission or  surface mass density. Similar results are obtained if we project the offset along the line that connects the galaxy to the peak of the X-ray emission. Nonetheless,  Spearman rank-order correlation tests show that these trends are not statistically significant. Likewise, it seems that the extent of the \Ha size with respect to the continuum size does not strongly correlate with the X-ray emission nor the surface mass density, as confirmed by a Spearman rank-order correlation test.

Overall, it seems that shocks and strong gradients in the X-ray IGM might alter the relative position between the peak of the \Ha emission and the peak of the light of the   recent star formation, even though we do not detect clear signs of gas  compression and/or stripping.

 We note that the lack of strong correlations does not allow us to identify a unique strong environmental effect that originates from the cluster center. We hypothesize that local effects, uncorrelated to the cluster-centric radius, play a larger role. Such effects weaken potential radial trends.

\subsection{Notes on remarkable objects}
In the following we describe some interesting objects presented in Figures 5-10.

Among galaxies with \Ha size similar to the rest-frame UV continuum size (Figures \ref{fig:c_size_Hec} and \ref{fig:f_size_Hec}), 
we note that \ma-00236 is a spiral galaxy with three main peaks of \Ha emission. Indeed, the strongest emission comes from the two spiral arms, while the flux in the core of the galaxy is less important. A complex \Ha structure extends throughout the entire galaxy. Only one PA covers the entire galaxy, while the other misses one of the  arms. 
In contrast, \ma-01477, even though showing a similar appearance in the color image to \ma-00236, is characterized by a much weaker and clumpy \Ha emission. \mb-01771 shows extended features both in the continuum and in the \Ha light. 

Among galaxies with \Ha size larger than the continuum size (Figures \ref{fig:c_size_Hgc} and \ref{fig:f_size_Hgc}), 
\ma-02189 and \ma-02574 show very elongated shapes in the continuum, but quite regular \Ha emission extended in both sizes. \mb-00229, \mb-01253 and \mb-01516 show very regular shapes in the continuum and very extended \Ha emission, in the case of \mb-01253 the emission is only in one direction. 
In \mb-00319 the \Ha emission is orthogonal to the continuum emission.  

Finally, among galaxies with \Ha size smaller than the continuum size (Figures \ref{fig:c_size_Hsc} and \ref{fig:f_size_Hsc}),
\ma-02334 shows an \Ha emission which is bent with respect to the continuum. In this case, the truncated \Ha disk might be an example of ram pressure stripping, which removed the ISM. The orientation of the tail does not point away from the cluster center, but the bending might suggests that the galaxy formed from an infalling population experiencing the cluster environment for the first time \citep[see, e.g.,][]{smith10}.

It is also worth noting that \ma-01305  (Fig. \ref{fig:c_size_Hec}) is very close to the cluster center, is located on a surface mass density distribution  peak and is quite close to a peak in the X-ray emission. The shape of the \Ha emission seems  to be unaffected by this peculiar position within the cluster, but the galaxy shows the largest projected offset along the line of sight.

\subsection{Star Formation Rates}

\begin{figure}
\centering
\includegraphics[scale=0.45]{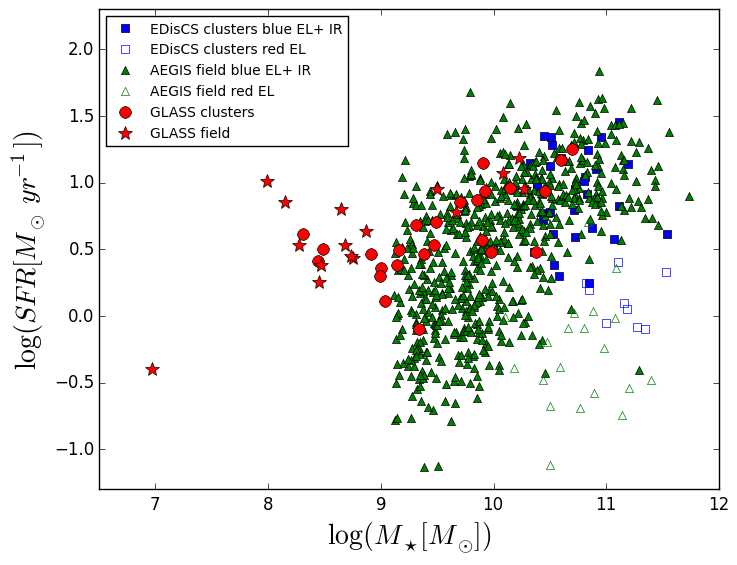}
\caption{GLASS cluster SFR-mass relation over plotted to the field relation at similar redshift \citep[from][]{noeske07} and the cluster relation at similar redshift  \citep[from][]{vulcani10}.  Red filled circles: GLASS clusters; red filled stars: GLASS field. Blue squares: EDisCS clusters, green triangles: AEGIS field. Empty symbols: red galaxies with detected emission lines, filled symbols: blue galaxies with detected emission lines and galaxies detected at 24$\mu m$ \citep[refer to][for details on the sample selection]{noeske07, vulcani10}. 
\label{fig:sfr_mass}}
\end{figure}

Figure \ref{fig:sfr_mass} shows the SFR-mass plane for our cluster and field galaxies, together with that found by \cite{noeske07} for field galaxies at $z\sim0.5$ and  by \cite{vulcani10} for cluster galaxies also at $z\sim0.5$. Our galaxies lay on the  field SFR-mass relation of blue galaxies with emission lines or detected in the Infrared \citep[see][for details on their sample selection]{noeske07, vulcani10}, and even trace the upper limit. To some extent, this was expected having selected star forming cluster galaxies. This results, however,  shows that at these redshifts cluster galaxies can be as star forming as field galaxies of similar mass.  The location of our galaxies on the plane is also in line with what was found by \cite{poggianti15} for the local universe, who showed that  galaxies with signs of ongoing stripping tend to be located above the best fit to the relation, indicating a SFR excess with respect to  galaxies  of the same mass but that are not being stripped. 
Recall that the field galaxies span a wide redshift range (0.2<$z$<0.7), therefore they lay on different regions of the SFR-mass plane simply due to the evolution of the SFR-M$_\ast$ relation with $z$ \citep[e.g.,][]{noeske07}.

Both in clusters and in the field, the $\Sigma$SFR ranges from $\sim$0.1 to 1 $M_\odot \, \textrm{yr}^{-1} \, \textrm{kpc}^{-2}$, suggesting that the physical conditions in star forming galaxies do not strongly depend on environment. 

\section{Discussion and Conclusions}
In this pilot study we have carried out an exploration of the spatial
distribution of star formation in galaxies beyond the local universe,
as traced by the \Ha\ emission in two of the GLASS clusters, \ma and
\mb. For this purpose, we have developed a new methodology to
produce \Ha maps taking advantage of the WFC3-G102 data at two
orthogonal PAs.  We then visually selected galaxies with \Ha emission
and, based on their redshift, assigned their membership to the
cluster. We have used galaxies in the foreground and background of the
two clusters to compile a field control sample. Both for field and
cluster galaxies, we computed the extent of the emission and its
position within the galaxy and compared these quantities to the
 younger stellar population  as traced by the rest frame UV continuum 
(obtained by images in the F475W filter) and the older stellar population as traced 
by the rest-frame optical continuum (obtained by images in the F110W filter).  We correlated galaxy properties
to global and local cluster properties, in order to look for signs of
cluster specific processes.

The main results of this analysis can be summarized as follows:

\begin{itemize}
\item Both in clusters and the field $\sim$60\% of the galaxies are more extended in \Ha than in the rest-frame UV continuum.  The emission appears larger in the cluster than in field galaxies. Trends are driven by a subpopulation ($\sim$20\%) of cluster galaxies with \Ha emission at least three time as extended the continuum emission. 

\item Both in clusters and in the field there is an offset between the peak of the \Ha emission and that in the rest-frame UV continuum. The displacement can reach $\sim$1 kpc. In clusters the offset appears to be marginally larger.

\item Comparing the extent of the offset and the cluster properties, we find a tentative correlation  between the projected offset and both the X-ray emission and the mass surface density: the larger the emission, the bigger the offset between the emission in the rest-frame UV continuum and the emission in \Ha. This offset seems to also to point at a cluster-specific mechanism.

\item \ma and \mb present very different surface mass density distribution and X ray emission, indicating that \mb  is much more relaxed than \ma, which in contrast presents a double peak in the distribution. In \mb all 
galaxies have \Ha disk larger than the rest-frame UV continuum, while in \ma galaxies with both extend and truncated \Ha are observed.
 This finding suggests that gradients in the X-ray IGM might alter the relative position between the peak of the \Ha emission and the peak of the light of the young stellar population, even though at this stage correlations are not supported by statistical tests. 
\end{itemize}

From our analysis a complex picture emerges and a simple explanation
can not describe our observations. Even though galaxies in clusters
and in the field present similar \Ha properties, the variety of their
morphology suggests that they are at different stages of their
evolution, therefore there can not be a unique mechanism acting on
galaxies in the different environments.  For example, the larger
extension of \Ha\ with respect to the continuum seems to be a generic
indication of inside-out growth \citep[see also][]{nelson12}. Specific
examples of this case include \mb-01253 and \mb-01910 in clusters
(Fig.\ref{fig:c_size_Hgc}) and \mb-01972 in the field
(Fig.\ref{fig:f_size_Hgc}).
Investigating whether this growth is localized in a disk component
will require careful bulge to disk decompositions which is planned for
a future work. However, given the variety of morphologies, well-ordered
disks do not appear to be the only site of star formation. 

Furthermore, the larger average sizes in the cluster point to an
additional cluster-specific mechanism responsible for stripping the
ionized gas or perhaps for triggering additional star formation in the
outskirts of the galaxies. The mechanism could be ram pressure
stripping of the ionized gas or perhaps tidal compression of the
outskirts or the galaxies, or both.
Galaxies \ma-02189 and \mb-00446 (Fig.\ref{fig:c_size_Hgc}) are
examples of this possible mechanism.

Finally, in some cases, galaxies have been already deprived from their
gas and are left with a smaller \Ha disk than the  recent star formation. 
\ma-02334 (Fig.\ref{fig:f_size_Hsc}) is a clear example of
galaxies in such stage.

 We note that the observed differences between cluster and field galaxies might be also due to the different mass and redshift distributions of the two samples, and not only to purely environmental effects.

The results from this pilot study illustrate the power and feasibility
of space-based grism data to learn qualitatively new information about
the mechanisms that regulate star formation in different environments
during the second half of the life of the universe. Having developed
the methods we are now in a position to carry out a larger scale
investigation on the full GLASS cluster sample, when visual
inspections and quality controls have been completed.  The extended
analysis will allow us not only to further distinguish and classify
the processes acting in clusters from those acting on the field, but
also to better correlate galaxy properties with the cluster global
properties, investigating in detail the role of the environment in
shutting down star formation.

\section*{Acknowledgments}

Support for GLASS (HST-GO-13459) was provided by NASA through a grant
from the Space Telescope Science Institute, which is operated by the
Association of Universities for Research in Astronomy, Inc., under
NASA contract NAS 5-26555. We are very grateful to the staff of the
Space Telescope for their assistance in planning, scheduling and
executing the observations.  B.V. acknowledges the support from the
World Premier International Research Center Initiative (WPI), MEXT,
Japan and the Kakenhi Grant-in-Aid for Young Scientists (B)(26870140)
from the Japan Society for the Promotion of Science (JSPS).

\bibliographystyle{apj}
\bibliography{biblio_SFR}

\label{lastpage}
\end{document}